\documentclass[12pt,prd,showpacs,tightenlines,nofootinbib]{revtex4}
\usepackage{bm}
\usepackage{graphics}
\usepackage{rotating}
\usepackage{epsfig}
\begin{document}
\title{\begin{flushright}{\rm\normalsize HU-EP-02/45}\end{flushright}
Properties of heavy quarkonia and $\bm{B_c}$ mesons in the
relativistic quark model}  
\author{D. Ebert}
\affiliation{Institut f\"ur Physik, Humboldt--Universit\"at zu Berlin,
Invalidenstrasse 110, D-10115 Berlin, Germany}
\author{R. N. Faustov}
\author{V. O. Galkin}
\affiliation{Institut f\"ur Physik, Humboldt--Universit\"at zu Berlin,
Invalidenstrasse 110, D-10115 Berlin, Germany}
\affiliation{Russian Academy of Sciences, Scientific Council for
Cybernetics, Vavilov Street 40, Moscow 117333, Russia}
\begin{abstract}
 The mass spectra and electromagnetic decay rates of charmonium,
 bottomonium and $B_c$ mesons are comprehensively investigated in the
 relativistic quark model. The presence of only heavy quarks allows the
 expansion in powers of their velocities. All relativistic
 corrections of order $v^2/c^2$, including retardation effects and
 one-loop radiative corrections, are systematically taken into account
 in the computations of the mass spectra. The obtained 
 wave functions are used for the 
 calculation of radiative magnetic dipole (M1) and electric dipole
 (E1) transitions. It is found that 
 relativistic effects play a substantial role. Their account and the
 proper choice of the Lorentz structure of the quark-antiquark
 interaction in a meson is crucial for bringing theoretical predictions
 in accord with experimental data. A detailed comparison of the
 calculated decay rates and branching fractions with available
 experimental data for radiative decays of 
 charmonium and bottomonium is presented. The possibilities to observe
 the currently missing spin-singlet $S$ and $P$ states as well as $D$
 states in bottomonium are discussed. The results for $B_c$ masses and
 decays are compared with other quark model predictions.   
\end{abstract}

\maketitle

\section{Introduction}
\label{sec:intro}

The investigation of the properties of mesons composed of a heavy quark and
antiquark ($b\bar b$, $c\bar c$, $c\bar b$)
gives very important insight into  heavy quark dynamics. Heavy
quarkonia have a rich spectroscopy with many narrow states lying under
the threshold of open flavour production. Excited states experience
different decays among which there are radiative transitions to lower
levels. The theoretical analysis shows that many properties of heavy
quarkonia, including mass spectra and radiative decay rates, are
significantly influenced by relativistic effects.  Thus their
inclusion is necessary for the correct description of the spectroscopy
and the determination of quarkonium wave functions. Radiative decays
are  the most sensitive to relativistic effects. Some of these decays,
which are forbidden in the exact nonrelativistic limit (so-called
hindered transitions) due to the orthogonality of initial and final
meson wave functions, have decay rates of the same order as the allowed
ones. In the relativistic description of mesons an important role
is played by properties of the confining quark-antiquark interaction,
in particular its Lorentz structure. Thus comparison of theoretical
predictions with experimental data can provide valuable information on
the form of the confining potential. Such information is of great
practical interest, since at present it is not possible to obtain the
$Q\bar Q$ potential in the whole  range of distances from the basic
principles of quantum chromodynamics (QCD). As it is well known, the
growing of the strong coupling 
constant with distance makes perturbation theory inapplicable at large
distances (in the infrared region). In this region it 
is necessary to account for nonperturbative effects connected with the
complicated structure of the QCD vacuum. All this leads to 
a theoretical uncertainty in the $Q\bar Q$ potential at large and
intermediate  distances. It is just in this region of large and
intermediate distances that most of the basic meson characteristics
are formed.     
   
At present,
a vast set of experimental data is available on the masses and
different decays of heavy quarkonia.  However,
not all states predicted by theory have been observed yet, while the others
need confirmation. Such missing or unconfirmed states are present both
in charmonium (${2\,}^1\!S_0$, ${1\,}^1\!P_1$, ${1\,}^1\!D_2$,
${1\,}^3\!D_{2,3}$) and bottomonium (spin-singlet ${}^1\!S_0$ and
${}^1\!P_1$ states,  $D$ states). The different possibilities for their 
experimental observation are proposed and widely discussed in the literature
\cite{elq,suz,kuang,grnb,gr,grD}. Radiative transitions from the
spin-triplet levels with $J^{PC}=1^{--}$ to these states as well as
their subsequent radiative decays play an important role in these
proposals.  
The missing charmonium states can also be searched in $B$ meson decays and
identified by their radiative transitions \cite{elq,suz}. For 
this purpose, reliable relativistic predictions for the masses of
these states and for the rates of radiative transitions involving them
are necessary.  

The properties of the $B_c$ meson are of special interest, since it
is the only heavy meson consisting of two heavy quarks with different
flavour. This difference of quark flavours forbids annihilation into
gluons. As a result, the excited $B_c$ meson states lying below the
$BD$ production threshold undergo pionic or radiative transitions to
the ground pseudoscalar state which then decays weakly. There should
be a rather rich set of such narrow states which are considerably more
stable than corresponding charmonium or bottomonium states. The CDF
collaboration \cite{cdf} reported the discovery of the $B_c$ ground
state in $p\bar p$ collisions at Fermilab. More experimental data are
expected to come in the near future from new hadronic colliders. 

The purpose of this paper is to give a detailed analysis of mass
spectra and radiative transitions in charmonium, bottomonium and $B_c$
mesons with the comprehensive account of the relativistic effects. This
will allow one to get valuable information about the Lorentz structure of
confining quark interactions from the comparison of obtained
predictions with available experimental data. On the other hand, it
will indicate the processes in which the missing states can be searched
for.          

The paper is organized as follows. In Sec.~\ref{rqm} we describe our
relativistic quark model. The expression for the heavy quark-antiquark
quasipotential with the account of relativistic 
(including retardation effects) and one loop radiative corrections is
given in Sec.~\ref{sec:bc}. There it is applied to the calculation of the
charmonium, bottomonium and $B_c$ meson mass spectra. In Sec.~\ref{sec:pvdc} 
pseudoscalar and vector decay constants of the $B_c$ meson are
calculated with the account of relativistic corrections and compared
with other theoretical predictions. In Sec.~\ref{rad} the relativistic
expressions for the radiative transition matrix elements in the
quasipotential approach are given. They are used for the
calculation of the  decay rates of radiative M1 and E1 transitions in
Secs.~\ref{m1} and \ref{e1}, respectively. The role of relativistic
effects in these transitions is investigated. Special attention is
payed to the
influence of the Lorentz structure of the quark potential  on the
relativistic corrections to decay rates. Pure vector and scalar
potentials as well as their mixture are considered. The obtained results are
compared with available experimental data, and the possibilities for
searching the missing states in bottomonium are discussed. Finally, our
conclusions are given in Sec.~\ref{sec:conc}.

\section{Relativistic quark model}  
\label{rqm}

In the quasipotential approach a meson is described by the wave
function of the bound quark-antiquark state, which satisfies the
quasipotential equation \cite{3} of the Schr\"odinger type \cite{4}
\begin{equation}
\label{quas}
{\left(\frac{b^2(M)}{2\mu_{R}}-\frac{{\bf
p}^2}{2\mu_{R}}\right)\Psi_{M}({\bf p})} =\int\frac{d^3 q}{(2\pi)^3}
 V({\bf p,q};M)\Psi_{M}({\bf q}),
\end{equation}
where the relativistic reduced mass is
\begin{equation}
\mu_{R}=\frac{E_1E_2}{E_1+E_2}=\frac{M^4-(m^2_1-m^2_2)^2}{4M^3},
\end{equation}
and $E_1$, $E_2$ are given by
\begin{equation}
\label{ee}
E_1=\frac{M^2-m_2^2+m_1^2}{2M}, \quad E_2=\frac{M^2-m_1^2+m_2^2}{2M}.
\end{equation}
Here $M=E_1+E_2$ is the meson mass, $m_{1,2}$ are the quark masses,
and ${\bf p}$ is their relative momentum.  
In the center of mass system the relative momentum squared on mass shell 
reads
\begin{equation}
{b^2(M) }
=\frac{[M^2-(m_1+m_2)^2][M^2-(m_1-m_2)^2]}{4M^2}.
\end{equation}

The kernel 
$V({\bf p,q};M)$ in Eq.~(\ref{quas}) is the quasipotential operator of
the quark-antiquark interaction. It is constructed with the help of the
off-mass-shell scattering amplitude, projected onto the positive
energy states. 
Constructing the quasipotential of the quark-antiquark interaction, 
we have assumed that the effective
interaction is the sum of the usual one-gluon exchange term with the mixture
of long-range vector and scalar linear confining potentials, where
the vector confining potential
contains the Pauli interaction. The quasipotential is then defined by
\cite{mass1}
  \begin{equation}
\label{qpot}
V({\bf p,q};M)=\bar{u}_1(p)\bar{u}_2(-p){\mathcal V}({\bf p}, {\bf
q};M)u_1(q)u_2(-q),
\end{equation}
with
$${\mathcal V}({\bf p},{\bf q};M)=\frac{4}{3}\alpha_sD_{ \mu\nu}({\bf
k})\gamma_1^{\mu}\gamma_2^{\nu}
+V^V_{\rm conf}({\bf k})\Gamma_1^{\mu}
\Gamma_{2;\mu}+V^S_{\rm conf}({\bf k}),$$
where $\alpha_S$ is the QCD coupling constant, $D_{\mu\nu}$ is the
gluon propagator in the Coulomb gauge
\begin{equation}
D^{00}({\bf k})=-\frac{4\pi}{{\bf k}^2}, \quad D^{ij}({\bf k})=
-\frac{4\pi}{k^2}\left(\delta^{ij}-\frac{k^ik^j}{{\bf k}^2}\right),
\quad D^{0i}=D^{i0}=0,
\end{equation}
and ${\bf k=p-q}$; $\gamma_{\mu}$ and $u(p)$ are 
the Dirac matrices and spinors
\begin{equation}
\label{spinor}
u^\lambda({p})=\sqrt{\frac{\epsilon(p)+m}{2\epsilon(p)}}
\left(
\begin{array}{c}1\cr {\displaystyle\frac{\bm{\sigma}
      {\bf  p}}{\epsilon(p)+m}}
\end{array}\right)\chi^\lambda,
\end{equation}
with $\epsilon(p)=\sqrt{p^2+m^2}$.
The effective long-range vector vertex is
given by
\begin{equation}
\label{kappa}
\Gamma_{\mu}({\bf k})=\gamma_{\mu}+
\frac{i\kappa}{2m}\sigma_{\mu\nu}k^{\nu},
\end{equation}
where $\kappa$ is the Pauli interaction constant characterizing the
anomalous chromomagnetic moment of quarks. Vector and
scalar confining potentials in the nonrelativistic limit reduce to
\begin{eqnarray}
\label{vlin}
V_V(r)&=&(1-\varepsilon)Ar+B,\nonumber\\ 
V_S(r)& =&\varepsilon Ar,
\end{eqnarray}
reproducing 
\begin{equation}
\label{nr}
V_{\rm conf}(r)=V_S(r)+V_V(r)=Ar+B,
\end{equation}
where $\varepsilon$ is the mixing coefficient. 

The expression for the quasipotential of the heavy quarkonia,
expanded in $v^2/c^2$ without and with retardation corrections to the
confining potential, can be found in Refs.~\cite{mass1} and \cite{mass},
respectively. The 
structure of the spin-dependent interaction is in agreement with
the parameterization of Eichten and Feinberg \cite{ef}. The
quasipotential for the heavy quark interaction with light antiquark
without employing the expansion in inverse powers of the light quark
mass is given in Ref.~\cite{hlm}.  
All the parameters of
our model such as quark masses, parameters of the linear confining potential
$A$ and $B$, mixing coefficient $\varepsilon$ and anomalous
chromomagnetic quark moment $\kappa$ are fixed from the analysis of
heavy quarkonium masses \cite{mass} (see  Sec.~\ref{sec:bc}) and radiative
decays \cite{gf}
(see Secs.~\ref{rad}--\ref{e1}). The quark masses
$m_b=4.88$ GeV, $m_c=1.55$ GeV and
the parameters of the linear potential $A=0.18$ GeV$^2$ and $B=-0.16$ GeV
have usual values of quark models.  The value of the mixing
coefficient of vector and scalar confining potentials $\varepsilon=-1$
has been determined from the consideration of the heavy quark expansion
for the semileptonic $B\to D$ decays
\cite{fg} and charmonium radiative decays \cite{gf}.
Finally, the universal Pauli interaction constant $\kappa=-1$ has been
fixed from the analysis of the fine splitting of heavy quarkonia ${
}^3P_J$- states \cite{mass1}. Note that the 
long-range  magnetic contribution to the potential in our model
is proportional to $(1+\kappa)$ and thus vanishes for the 
chosen value of $\kappa=-1$. In  the present paper we will take 
into consideration the retardation corrections as well as one-loop  
radiative corrections.

\section{Heavy quarkonium and $\bm{B_{\lowercase{c}}}$ meson mass spectra}
\label{sec:bc}

The heavy quark-antiquark potential with the account of retardation
effects and one loop radiative corrections can be 
presented in the form of a sum of  spin-independent and spin-dependent
parts. For the spin-independent part we have \cite{mass} 
\begin{eqnarray}
\label{sipot}
V_{\rm SI}(r)&=&-\frac43\frac{\bar \alpha_V(\mu^2)}{r} +Ar+B -\frac43
\frac{\beta_0 \alpha_s^2(\mu^2)}{2\pi}\frac{\ln(\mu r)}{r} \cr\nonumber\\
&& +\frac18\left(\frac{1}{m_1^2}+\frac{1}{m_2^2}\right) 
\Delta\left[ -\frac43\frac{\bar 
\alpha_V(\mu^2)}{r} -\frac43\frac{\beta_0\alpha_s^2(\mu^2)}{2\pi}
\frac{\ln(\mu r)}{r}
 +(1-\varepsilon)(1+2\kappa)Ar\right]\cr\nonumber\\
&&+\frac{1}{2m_1m_2}\Biggl(\left\{-\frac43\frac{\bar\alpha_V}{r}
\left[{\bf p}^2+\frac{({\bf p\cdot r})^2}{r^2}\right]\right\}_W\cr\nonumber\\
&&-\frac43\frac{\beta_0\alpha_s^2(\mu^2)}{2\pi}\left\{{\bf p}^2
  \frac{\ln(\mu r)}{r} +\frac{({\bf p\cdot 
r})^2}{r^2}\left(\frac{\ln(\mu r)}{r}-\frac1r\right)\right\}_W\Biggr)
\cr\nonumber\\
&& +\left[\frac{1-\varepsilon}{2m_1m_2}-\frac{\varepsilon}{4}
\left(\frac{1}{m_1^2}+\frac{1}{m_2^2}\right)\right]\left\{Ar\left[{\bf p}^2
-\frac{({\bf p\cdot r})^2}{r^2}\right]\right\}_W\cr\nonumber\\
&&+\left[\frac{1}{4}\left(\frac{1}{m_1^2}+\frac{1}{m_2^2}\right)+
\frac{1}{m_1m_2}\right]B{\bf p}^2,
\end{eqnarray}
where
\begin{eqnarray}
\bar\alpha_V(\mu^2)&=&\alpha_s(\mu^2)\left[1+\left(\frac{a_1}{4}
+\frac{\gamma_E\beta_0}{2}\right)\frac{\alpha_s(\mu^2)}{\pi}\right],\\
a_1&=&\frac{31}{3}-\frac{10}{9}n_f,\cr
\beta_0&=&11-\frac23n_f.\nonumber
\end{eqnarray}
Here $n_f$ is the number of flavours and $\mu$ is a renormalization
scale. Note that for a quantity quadratic in the momenta we use the
Weyl prescription \cite{bcp}:
$$\{f(r)p^ip^j\}_W=\frac14\{\{f(r),p^i\},p^j\}.$$

For the dependence of the QCD coupling constant $\alpha_s(\mu^2)$ 
on the renormalization point $\mu^2$ we use the leading order
result
\begin{equation}
\label{alpha}
\alpha_s(\mu^2)=\frac{4\pi}{\beta_0\ln(\mu^2/\Lambda^2)}.
\end{equation}
In our numerical calculations we set the renormalization scale 
$\mu=2m_1m_2/(m_1+m_2)$ and $\Lambda=0.168$~GeV, which gives
$\alpha_s=0.314$ for $m_1=m_2=m_c$ (charmonium); $\alpha_s=0.223$ for
$m_1=m_2=m_b$ (bottomonium); and $\alpha_s=0.265$ for $m_1=m_c$, $m_2=m_b$
($B_c$ meson).

The spin-dependent part of the quark-antiquark potential for unequal
quark masses  with the inclusion of radiative corrections \cite{gupta,ptn} 
can be presented in our model as follows:
\begin{equation}
\label{vsd}
V_{\rm SD}(r)= a\ {\bf L}\cdot{\bf S}+b\left[\frac{3}{r^2}({\bf S}_1\cdot
{\bf r})({\bf S}_2\cdot {\bf r})-({\bf S}_1\cdot {\bf S}_2)\right] +c\ {\bf
S}_1\cdot {\bf S}_2 +d\ {\bf L}\cdot({\bf S}_1-{\bf S}_2), 
\end{equation}
\begin{eqnarray}
\label{a}
a&=& \frac14\left(\frac{1}{m_1^2}+\frac{1}{m_2^2}\right)\Biggl\{
  \frac43\frac{\alpha_s(\mu^2)}{r^3}\Biggl(1+ 
\frac{\alpha_s(\mu^2)}{\pi}\Biggl[\frac73-\frac{\beta_0}{12}+
  \gamma_E\left(\frac{\beta_0}{2}-3\right)+\frac{\beta_0}{2}\ln(\mu r)
\cr\nonumber\\ 
&&-3\ln(\sqrt{m_1m_2}\,r)\Biggr]\Biggr)
-\frac{A}{r}\Biggr\}+\frac1{m_1m_2}\frac43\frac{\alpha_s(\mu^2)}{r^3}
\Biggl(1+ 
\frac{\alpha_s(\mu^2)}{\pi}\Biggl[\frac16-\frac{\beta_0}{12}\cr\nonumber\\
&&+\gamma_E\left(\frac{\beta_0}{2}-\frac32\right)+\frac{\beta_0}{2}
\ln(\mu r)
-\frac32\ln(\sqrt{m_1m_2}\,r)\Biggr]\Biggr)\cr\nonumber\\
&&+\left(\frac{1}{m_1^2}-\frac{1}{m_2^2}\right)
  \frac{\alpha_s^2(\mu^2)}{2\pi r^3}\ln\frac{m_2}{m_1}
  +\frac12\left(\frac{1}{m_1}+\frac{1}{m_2}\right)^2 
  (1+\kappa)(1-\varepsilon)\frac{A}{r},\\ \cr
\label{b}
b&=& \frac{1}{3m_1 m_2}\Biggl\{\frac{4\alpha_s(\mu^2)}{r^3}\Biggl(1+
\frac{\alpha_s(\mu^2)}{\pi}\Biggl[\frac{29}{6}-\frac{1}{4}\beta_0+
\gamma_E\left(\frac{\beta_0}{2}-3\right)+\frac{\beta_0}{2}\ln(\mu r),
\cr\nonumber\\
&&-3\ln(\sqrt{m_1m_2}\,r)\Biggr]\Biggr)
+(1+\kappa)^2(1-\varepsilon)\frac{A}{r}\Biggr\},\\ \cr
\label{c}
c&=& \frac{4}{3m_1 m_2}\Biggl\{\frac{8\pi\alpha_s(\mu^2)}{3}\Biggl(\left[1+
\frac{\alpha_s(\mu^2)}{\pi}\left(\frac{5}{12}\beta_0-\frac{11}{3} 
-\left[\frac{m_1-m_2}{m_1+m_2}+\frac18\,\frac{m_1+m_2}{m_1-m_2}\right]
\ln\frac{m_2}{m_1}\right)
\right]\delta^3(r)\cr\nonumber\\
&&+\frac{\alpha_s(\mu^2)}{\pi}\left[-
\frac{\beta_0}{8\pi}\nabla^2\left(\frac{\ln({\mu} r)+\gamma_E}{r}\right)
+\frac{21}{16\pi}\nabla^2\left(
\frac{\ln(\sqrt{m_1m_2}\,r)+\gamma_E}{r}\right)\right]\Biggr)\cr\nonumber\\
&&+(1+\kappa)^2(1-\varepsilon)\frac{A}{r}\Biggr\},\\ \cr
\label{d}
d&=& \frac14\left(\frac{1}{m_1^2}-\frac{1}{m_2^2}\right)\Biggl\{
  \frac43\frac{\alpha_s(\mu^2)}{r^3}\Biggl(1+ 
\frac{\alpha_s(\mu^2)}{\pi}\Biggl[\frac73-\frac{\beta_0}{12}+
  \gamma_E\left(\frac{\beta_0}{2}-3\right)+\frac{\beta_0}{2}\ln(\mu r)
\cr\nonumber\\
&&-3\ln(\sqrt{m_1m_2}\,r)\Biggr]\Biggr)
-\frac{A}{r}-2(1+\kappa)(1-\varepsilon)\frac{A}{r}\Biggr\}
+\left(\frac{1}{m_1}+\frac{1}{m_2}\right)^2
  \frac{\alpha_s^2(\mu^2)}{2\pi r^3}\ln\frac{m_2}{m_1},
   \end{eqnarray}
where ${\bf L}$ is the orbital momentum and ${\bf S}_{1,2}$, ${\bf
S}={\bf S}_1+ {\bf S}_2$ are the spin momenta. For the equal mass 
case ($m_1=m_2=m$) the second order in $\alpha_s$ contribution of the
annihilation diagrams  
\begin{equation}
  \label{eq:deltc}
  \delta c=\frac{8\alpha_s^2(\mu^2)}{3m^2}\left(1-\ln
  2\right)\delta^3(r) 
\end{equation}
must be added to the spin-spin interaction coefficient $c$ in
Eq.~(\ref{c}). 

The correct description of the fine structure of the heavy quarkonium
mass spectrum requires the vanishing of the vector confinement contribution.
This can be achieved by setting $1+\kappa=0$, i.e., the total
long-range quark chromomagnetic moment equals zero, which is in accord
with the flux tube \cite{buch} and minimal area \cite{bcp,bv} models.
One can see from Eq.~(\ref{vsd}) that for the spin-dependent part of
the potential this conjecture is equivalent to
the assumption about the scalar structure of confining interaction
\cite{scal}. 

To calculate the heavy meson mass spectra  with the account
of all relativistic corrections (including retardation effects) of
order $v^2/c^2$  and one-loop radiative corrections we substitute the
quasipotential which is a sum of the spin-independent  (\ref{sipot})
and spin-dependent (\ref{vsd}) parts into the quasipotential equation
(\ref{quas}). Then we multiply the resulting expression from the left
by the quasipotential wave function of a bound state and integrate
with respect to the relative momentum. Taking into account the
accuracy of the calculations, we can use for the resulting matrix
elements the wave functions of Eq.~(\ref{quas}) with the static
potential \footnote{This static potential includes also some
radiative corrections.  The remaining radiative correction
term with logarithm in Eq.~(\ref{sipot}), also not vanishing in the static
limit, is treated perturbatively.}  
\begin{equation}
V_{\rm NR}(r)=-\frac43\frac{\bar \alpha_V(\mu^2)}{r} +Ar+B.
\end{equation}
As a result we obtain the mass formula 
\begin{equation}
\label{mform}
\frac{b^2(M)}{2\mu_R}=W+\langle a\rangle\langle{\bf L}\cdot{\bf S}\rangle
+\langle b\rangle \langle\left[\frac{3}{r^2}
({\bf S}_1\cdot {\bf r})
({\bf S}_2\cdot {\bf r})-({\bf S}_1\cdot {\bf S}_2)\right] \rangle
+\langle c\rangle \langle{\bf S}_1\cdot {\bf S}_2\rangle + d \langle
{\bf L}\cdot({\bf S}_1-{\bf S}_2)\rangle ,
\end{equation}
where
\begin{equation}
W=\langle V_{\rm SI}\rangle+\frac{\langle {\bf p}^2\rangle}{2\mu_R}.
\end{equation}
The first term on the right-hand side of the mass formula
(\ref{mform}) contains all spin-independent contributions, the second
and the last terms describe the spin-orbit interaction, the third term is
responsible for the tensor interaction, while the forth term gives the
spin-spin interaction. The last term is not zero only for the unequal
mass case $m_1\ne m_2$ and leads to the mixing of triplet and singlet
meson states with the total angular momentum $J$ equal to the orbital
momentum $L$.    

In Table~\ref{charm} the calculated charmonium mass spectrum is
compared with experimental data. 
For meson states we use the notation $n\,{}^{2S+1}\!L_J$, where $n=n_r+1$ 
and $n_r$ is the radial quantum number. 
Our predictions agree
with PDG \cite{pdg} data within a few MeV. Our model correctly reproduces
both the position of the levels and their fine and hyperfine
splittings. In this Table we also give the recent Belle Collaboration
data \cite{belle} on pseudoscalar $\eta_c(1\,{}^1\!S_0)$ and
$\eta'_c(2\,{}^1\!S_0)$ states observed in $B$ decays. The measured
mass of the ground spin-singlet state  $\eta_c(1\,{}^1\!S_0)$ is in
good agreement with world averages and predictions of our model, while
the radially excited $\eta'_c(2\,{}^1\!S_0)$ state lies considerably
higher than previous experimental indications and most of the
theoretical predictions. If these data are confirmed, it will be
difficult to accommodate such a small hyperfine splitting $\approx
32$~MeV (almost four times smaller than 117~MeV splitting
for the ground state) in the framework of the quark
model.~\footnote{The position of the $\Psi'(2\,{}^3\! S_1)$ can  in
principle be influenced by the nearby threshold of the open charm
production.} 

\begin{table}[htb]
\caption{Charmonium mass spectrum (in GeV). }
\label{charm}
\begin{ruledtabular}
\begin{tabular}{ccccc}
State &Particle & Theory & \multicolumn{2}{c}{Experiment}  \\
$n\,{}^{2S+1}\! L_J$&  &  & PDG \cite{pdg} &
Belle \cite{belle} \\
\hline
$1\,{}^1\! S_0$& $\eta_c$ & 2.979 & 2.9797 &2.979\\
$1\,{}^3\! S_1$& $J/\Psi$ & 3.096 & 3.09687 &\\
$1\,{}^3\! P_0$& $\chi_{c0}$ & 3.424 & 3.4151& \\
$1\,{}^3\! P_1$& $\chi_{c1}$ & 3.510 & 3.51051 &\\
$1\,{}^3\! P_2$& $\chi_{c2}$ & 3.556 & 3.55618 &\\
$1\,{}^1\! P_1$& $h_c$ & 3.526& &\\
$2\,{}^1\! S_0$& $\eta_c'$ & 3.588 & 3.594\footnote{This value from
  Ref.~\cite{cb}  is included only in the PDG
  Listings.} & 3.654\\
$2\,{}^3\! S_1$& $\Psi'$     & 3.686 & 3.68596 &\\ 
$1\,{}^3\! D_1$&  & 3.798 & 3.7699\footnotemark[2] &\\
$1\,{}^3\! D_2$&  & 3.813 & &\\
$1\,{}^3\! D_3$&  & 3.815 & &\\
$1\,{}^1\! D_2$& & 3.811&&\\
$2\,{}^3\! P_0$& $\chi'_{c0}$ & 3.854 & &\\
$2\,{}^3\! P_1$& $\chi'_{c1}$ & 3.929 & &\\
$2\,{}^3\! P_2$& $\chi'_{c2}$ & 3.972 & &\\
$2\,{}^1\! P_1$& $h_c'$ & 3.945&&\\
$3\,{}^1\! S_0$& $\eta_c''$ & 3.991 & &\\
$3\,{}^3\! S_1$& $\Psi''$     & 4.088 & 4.040\footnotemark[2] &

\footnotetext[2]{ Mixture of $S$ and $D$ states.}
\end{tabular}
\end{ruledtabular}
\end{table} 

Our prediction for the mass of $h_c(1\,{}^1\! P_1)$ is consistent with
the data from the Fermilab Experiment E760 \cite{E760} on $p\bar p \to
h_c(3526) \to \pi^0 J/\Psi$ which, however, need confirmation. The same
is true for the indication of a $1\,{}^3\! D_2$ state with mass
$3836\pm 13$~MeV in $\pi^\pm N\to J/\Psi \pi^+\pi^-+$anything
\cite{E705}. 

For the calculation of the bottomonium mass spectrum it is also
necessary to take into account 
additional one-loop corrections due to the finite mass of the charm quark
\cite{hm,h,m,bsv}. We considered these corrections within our model in
Ref.~\cite{efgmc} and found that they give contributions of a few MeV
and are weakly dependent on the quantum numbers of the bottomonium
states. The one-loop correction to the static $Q\bar Q$ potential in QCD
due to the finite $c$ quark mass is given by \cite{m,efgmc}
\begin{equation}
  \label{eq:deltav}
  \Delta V(r,m_c)= -\frac49\frac{\alpha_s^2(\mu)}{\pi
  r}\left[\ln(\sqrt{a_0}m_c r) +\gamma_E+E_1(\sqrt{a_0}m_c
  r)\right],
\end{equation}  
where
\[
E_1(x)=\int_x^\infty e^{-t}\ \frac{dt}t=-\gamma_E-\ln x-
\sum_{n=1}^\infty \frac{(-x)^n}{n\cdot n!},
\] 
$\gamma_E\cong0.5772$ is the Euler constant  and $a_0=5.2$.   The
resulting bottomonium mass spectrum 
with the account of this correction is
given in Table~\ref{bottom}. We found that the small shift of the
QCD parameter $\Lambda$ from our previous \cite{mass} value 0.178
to 0.168~GeV (with all other parameters remaining fixed) allows us to get
a good fit to 
the bottomonium mass spectrum with the account of finite charm quark mass
corrections.  The difference
between the theoretical and experimental data is less than 3~MeV. Very
recently CLEO Collaboration presented \cite{cleo} the first evidence for the
production of the triplet $\Upsilon(1D)$ state in the four photon
cascades starting from $\Upsilon(3S)$. In Table~\ref{bottom} we give
their preliminary result for the mass of $\Upsilon(1D_2)$ state which is
consistent with our prediction.   

\begin{table}
\caption{Bottomonium mass spectrum (in GeV). }
\label{bottom}
\begin{ruledtabular}
\begin{tabular}{ccccc}
State & Particle &{Theory} & 
\multicolumn{2}{c}{Experiment} \\
$n\,{}^{2S+1}\! L_J$&  &  & PDG \cite{pdg}
&CLEO \cite{cleo} \\
\hline
$1\,{}^1\! S_0$& $\eta_b$  &9.400 &  &\\
$1\,{}^3\! S_1$& $\Upsilon$ & 9.460 & 9.46030 &\\
$1\,{}^3\! P_0$& $\chi_{b0}$ & 9.863 & 9.8599 & \\
$1\,{}^3\! P_1$& $\chi_{b1}$ & 9.892 & 9.8927 & \\
$1\,{}^3\! P_2$& $\chi_{b2}$ & 9.913 & 9.9126 &  \\
$1\,{}^1\! P_1$& $h_b$& 9.901&&\\
$2\,{}^1\! S_0$& $\eta_b'$ & 9.993 & &\\
$2\,{}^3\! S_1$& $\Upsilon'$  & 10.023 & 10.02326 &\\ 
$1\,{}^3\! D_1$&  & 10.153 &  & \\
$1\,{}^3\! D_2$&  & 10.158 &  &10.162\\
$1\,{}^3\! D_3$&  & 10.162 & &\\
$1\,{}^1\! D_2$&  & 10.158 &&\\
$2\,{}^3\! P_0$& $\chi'_{b0}$ & 10.234 & 10.2321 & \\
$2\,{}^3\! P_1$& $\chi'_{b1}$ & 10.255 & 10.2552 & 10.2556 \\
$2\,{}^3\! P_2$& $\chi'_{b2}$ & 10.268 & 10.2685 & 10.2688\\
$2\,{}^1\! P_1$& $h_b'$ & 10.261&&\\
$3\,{}^1\! S_0$& $\eta_b''$ & 10.328 & & \\
$3\,{}^3\! S_1$& $\Upsilon''$ & 10.355 & 10.3552 &\\
\end{tabular}
\end{ruledtabular}
\end{table}

For the mesons consisting of quarks with different flavours
($m_1\ne m_2$), such as the $B_c$ meson, the coefficient $d$ (\ref{d})
in the spin-dependent part of the quark potential (\ref{vsd}) is not equal
to zero. This results 
in the mixing of singlet and triplet $P$ states with $J=1$, 
\begin{eqnarray}
  \label{eq:pmix}
  nP1'&=&{n}\,{}^1\! P_1 \cos\theta_{nP}+{n}\,{}^3\! P_1 \sin\theta_{nP},\cr\cr
 nP1&=&-{n}\,{}^1\! P_1 \sin\theta_{nP}+{n}\,{}^3\! P_1 \cos\theta_{nP},
\end{eqnarray}
and of $D$ states with $J=2$,
\begin{eqnarray}
  \label{eq:dmix}
  nD2'&=&{n}\,{}^1\! D_2 \cos\theta_{nD}+{n}\,{}^3\! D_2 \sin\theta_{nD},\cr\cr
 nD2&=&-{n}\,{}^1\! D_2 \sin\theta_{nD}+{n}\,{}^3\! D_2 \cos\theta_{nD}.
\end{eqnarray}
For the $B_c$ meson the values of the mixing angles in our model are
\begin{equation}
  \label{eq:mixa}
  \theta_{1P}=0.357, \qquad \theta_{2P}=0.405, \qquad
  \theta_{1D}=-0.627. 
\end{equation}

\begin{table}[htbp]
  \caption{$B_c$ meson mass spectrum (in GeV).}
  \label{tab:bcm}
  \begin{ruledtabular}
  \begin{tabular}{cccccc}
State& & & & \\
$n\,{}^{2S+1}\! L_J$ & Our & \cite{eq} & \cite{gklt} &\cite{ful} &\cite{nl}\\
\hline
$1\,{}^1\! S_0$& 6.270 & 6.264 & 6.253 & 6.286& $\ge 6.2196$\\\
$1\,{}^3\! S_1$& 6.332 & 6.337 & 6.317 & 6.341& $\ge6.2786$ \\
$1\,{}^3\! P_0$& 6.699 & 6.700 & 6.683 & 6.701& $\ge6.6386$\\
$1P1$ & 6.734 & 6.730 & 6.717 & 6.737 &$\ge6.7012$\\
$1P1^{'}$&6.749& 6.736 & 6.729 & 6.760&$\ge6.7012$ \\
$1\,{}^3\! P_2$& 6.762 & 6.747 & 6.743 & 6.772 &$\ge6.7347$\\
$2\,{}^1\! S_0$& 6.835 & 6.856 & 6.867 & 6.882& \\
$2\,{}^3\! S_1$& 6.881 & 6.899 & 6.902 & 6.914& \\
$1\,{}^3\! D_1$& 7.072 & 7.012 & 7.008 & 7.019& \\
$1D2$ & 7.077 & 7.009 & 7.001 & 7.028& \\
$1D2^{'}$&7.079& 7.012 & 7.016 & 7.028& \\
$1\,{}^3\! D_3$& 7.081 & 7.005 & 7.007 & 7.032& \\
$2\,{}^3\! P_0$& 7.091 & 7.108 & 7.088 &    &  \\
$2P1$ & 7.126 & 7.135 & 7.113 &    &  \\
$2P1^{'}$&7.145& 7.142 & 7.124 &   &   \\
$2\,{}^3\! P_2$& 7.156 & 7.153 & 7.134 &    &  \\
$3\,{}^1\! S_0$& 7.193 & 7.244 &       &    &  \\
$3\,{}^3\! S_1$& 7.235 & 7.280 &       &    &  
  \end{tabular}
  \end{ruledtabular} 
\end{table}

In Table~\ref{tab:bcm} we compare our model predictions for the mass
spectrum of the $B_c$ meson with other quark model results
\cite{eq,gklt,ful,nl}. We see that the differences between the
predictions in most cases do not exceed 30 MeV. The only exceptions
are masses of $1D$ states, which are $50-70$ MeV heavier in our
model. The fine and hyperfine splittings are also consistent with each
other. All these predictions for the ground state pseudoscalar $B_c$
and vector $B_c^*$ meson masses satisfy the bounds found by Kwong and
Rosner \cite{kr}:
\[ 6.194~{\rm GeV}< M_{B_c}<6.292~{\rm GeV}\]
and
\[ 6.284~{\rm GeV}< M_{B_c^*}<6.357~{\rm GeV}.\]
In Ref.~\cite{bvbc} the ground state $B_c$ mass was evaluated in
perturbative QCD.
Experimental data \cite{pdg} at present are available only for the
$B_c$ ground state and have large error bars $M_{B_c}=6.4\pm 0.4$~GeV.

In the following sections we apply the masses and wave functions
of $\Psi$, $\Upsilon$ and $B_c$ mesons for the calculation of their
decay constants and decay rates.

\section{Pseudoscalar and vector decay constants of
the  $\bm{B_{\lowercase{c}}}$ meson}
\label{sec:pvdc}

The $B_c$ meson and its first excitations which lie below the $BD$
threshold are stable against strong decays, since they cannot annihilate
into gluons.  They can decay via electromagnetic and pionic
transitions into the lightest pseudoscalar ground state $B_c$. The
significant contribution to the $B_c$ total decay rate comes from the
annihilation of the $c$ quark and $\bar b$ antiquark into the vector
boson $W^+$ which decays into a lepton and a neutrino or a
quark-antiquark pair. The weak annihilation decay rate is determined by the
pseudoscalar constant of the $B_c$ meson.  

The decay constants $f_P$ and $f_V$ of the pseudoscalar ($P$) and
vector ($V$) mesons parameterize the matrix elements of the weak
current between the corresponding meson and the vacuum. In the case of
the $B_c$ meson they are defined by 
\begin{eqnarray}
  \label{eq:dc}
  \left<0|\bar b \gamma^\mu\gamma_5 c|P({\bf K})\right>&=& i f_P
  K^\mu,\\ 
\left<0|\bar b \gamma^\mu c|V({\bf K},\varepsilon)\right>&=& f_V
  M_V \varepsilon^\mu,
\end{eqnarray}
where ${\bf K}$ is the meson momentum, $P$ corresponds to the pseudoscalar
$B_c$ and $V$ to the vector $B_c^*$ mesons, $\varepsilon^\mu$ and $M_V$ are
the polarisation vector and mass of the vector $B_c^*$ meson.

In the relativistic quark model the decay constants can be expressed
through the meson wave function $\Phi_{P,V}(p)$ in the momentum space
and are given by \cite{gfm}
\begin{eqnarray}
  \label{eq:fpv}
  f_{P,V}&=&\sqrt{\frac{12}{M_{P,V}}}\int \frac{d^3
  p}{(2\pi)^3}\left(\frac{\epsilon_c(p)+m_c}{2\epsilon_c(p)}\right)^{1/2}
  \left(\frac{\epsilon_b(p)+m_b}{2\epsilon_b(p)}\right)^{1/2}
  \nonumber\\ \cr
&&\times \left\{ 1
  +\lambda_{P,V}\,\frac{{\bf p}^2}{[\epsilon_c(p)+m_c]
  [\epsilon_b(p)+m_b]}\right\}  \Phi_{P,V}(p),
\end{eqnarray}
with $\lambda_P=-1$ and $\lambda_V=1/3$.
In the nonrelativistic limit $p^2/m^2\to 0$ these expressions for
decay constants give the well-known formula
\begin{equation}
\label{eq:fnr}
f_P^{\rm NR}=f_V^{\rm NR}=
\sqrt{\frac{12}{M_{P,V}}}\left|\Psi_{P,V}(0)\right|,
\end{equation}
where $\Psi_{P,V}(0)$ is the meson wave function at the origin
$r=0$.

The calculated values of the pseudoscalar
and vector decay constants of the $B_c$ meson in our model using the
relativistic formula (\ref{eq:fpv}) are displayed in
Table~\ref{tab:fpv}. They are compared with the ones calculated using
the nonrelativistic expression (\ref{eq:fnr}) and  other predictions of
the nonrelativistic quark models  \cite{eq,ful}, QCD sum rules
\cite{gklt} and lattice NRQCD \cite{jw}. We see that inclusion of
relativistic corrections reduces the pseudoscalar decay constant
$f_{B_c}$ by 20\% and produces the difference between vector and
pseudoscalar decay constants of approximately 70 MeV. The calculated
values of these decay constants are consistent with lattice \cite{jw}
and QCD sum rule \cite{gklt} predictions. 

\begin{table}[htbp]
 \caption{Pseudoscalar and vector decay constants ($f_P=f_{B_c}$,
   $f_V=f_{B_c^*}$)  of the $B_c$ meson (in MeV).}
  \label{tab:fpv}
\begin{ruledtabular}
  \begin{tabular}{ccccccc}
Constant& rel & NR & \cite{eq} &\cite{gklt} &\cite{ful}&\cite{jw}\\
\hline
$f_{B_c}$& 433 & 562 & 500 & $460\pm60$ & 517 & $420\pm13$\\
$f_{B^*_c}$& 503 & 562 & 500 & $460\pm60$ & 517 & 
  \end{tabular}
 \end{ruledtabular}
\end{table}

\section{Radiative transitions in heavy quarkonia and
  $\bm{B_{\lowercase{c}}}$ mesons}
\label{rad}

\begin{figure}
  \centering
  \includegraphics{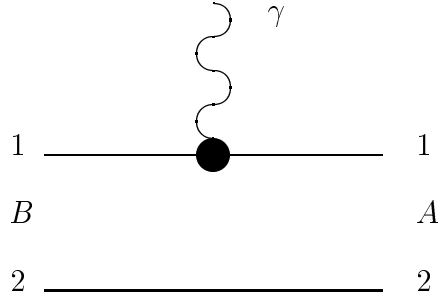}
  \caption{Lowest order vertex function $\Gamma^{(1)}$
corresponding to Eq.~(\ref{gam1}). Radiation only from one quark is shown.}
  \label{fig:1}
\end{figure}
\begin{figure}
  \centering
  \includegraphics{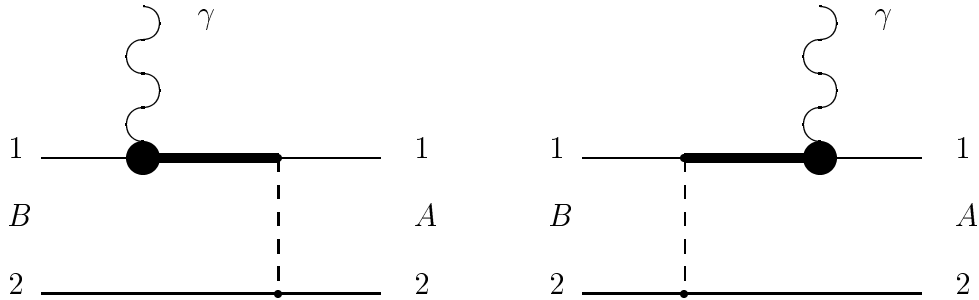}
  \caption{ Vertex function $\Gamma^{(2)}$
corresponding to Eq.~(\ref{gam2}). Dashed lines represent the interaction 
operator ${\mathcal V}$ in Eq.~(\ref{qpot}). Bold lines denote the  
negative-energy part of the quark propagator. As on Fig.~\ref{fig:1},
radiation only from one quark is shown.}
  \label{fig:2}
\end{figure}

To determine the rates of radiative decays ($B\to A+\gamma$) it is
necessary to calculate the matrix element of the
electromagnetic current $J_\mu$ between the initial ($B$) and final
($A$) meson states.
In the quasipotential approach such  matrix element has the form \cite{f}
\begin{equation}
\label{mxet}
\langle A \vert J_\mu (0) \vert B\rangle
=\int \frac{d^3p\, d^3q}{(2\pi )^6} \bar \Psi_{A\, {\bf P}}({\bf
p})\Gamma _\mu ({\bf p},{\bf q})\Psi_{B\, {\bf Q}}({\bf q}),
\end{equation}
where $\Gamma _\mu ({\bf p},{\bf
q})$ is the two-particle vertex function and  $\Psi_{A,B}$ are the
meson wave functions projected onto the positive energy states of
quarks and boosted to the moving reference frame.
The contributions to $\Gamma$ come from Figs.~1 and 2.
The contribution $\Gamma^{(2)}$ is the consequence of the
projection onto the positive-energy states. Note that the form
of the relativistic corrections resulting from the vertex function
$\Gamma^{(2)}$ explicitly depends on the Lorentz structure of the
$Q\bar Q$-interaction.  Thus the vertex function is given by
\begin{equation}
  \label{eq:gam}
  \Gamma_\mu({\bf p},{\bf q})=\Gamma_\mu^{(1)}({\bf p},{\bf q})+
  \Gamma_\mu^{(2)}({\bf p},{\bf q})+ \cdots ,
\end{equation}
where
\begin{equation}\label{gam1}
\Gamma_\mu ^{(1)}({\bf p},{\bf q})=e_1\bar
u_1(p_1)\gamma_\mu  
u_1(q_1)(2\pi)^3\delta({\bf p}_2-{\bf q}_2) +(1\leftrightarrow 2),
\end{equation}
and
\begin{eqnarray}\label{gam2}
\Gamma_\mu^{(2)}({\bf p},{\bf q})&=&e_1\bar u_1(p_1)\bar
u_2(p_2) \biggl\{{\mathcal V}({\bf p}_2-{\bf q}_2)
\frac{\Lambda_1^{(-)}(k_1')}{
\epsilon_1(k_1')+ \epsilon_1(q_1)}\gamma_1^0\gamma_{1\mu}
\nonumber\\
& & + \gamma_{1\mu}
\frac{\Lambda_1^{(-)}({k}_1)}{ \epsilon_1(k_1)+\epsilon_1(p_1)}
\gamma_1^0{\mathcal V}({\bf p}_2-{\bf q}_2)
\biggr\}u_1(q_1) u_2(q_2) +(1\leftrightarrow 2). 
\end{eqnarray}
Here $e_{1,2}$ are the quark charges, ${\bf k}_1={\bf p}_1-{\bf\Delta};
\quad {\bf k}_1'={\bf 
q}_1+{\bf\Delta};\quad {\bf\Delta}={\bf P}-{\bf Q}$; 
\[
\Lambda^{(-)}(p)={\epsilon(p)-\bigl( m\gamma ^0+\gamma^0(
\bm{\gamma}\cdot{\bf p})\bigr) \over 2\epsilon (p)}, \qquad \epsilon(p)=
\sqrt{p^2+m^2},
\]
and  
\begin{eqnarray*} 
p_{1,2}&=&\epsilon_{1,2}(p)\frac{p_{A}}{M_{A}}
\pm\sum_{i=1}^3 n^{(i)}(p_{A})p^i,\\
q_{1,2}&=&\epsilon_{1,2}(q)\frac{p_B}{M_B} \pm \sum_{i=1}^3 n^{(i)}
(p_B)q^i,\end{eqnarray*}
where $n^{(i)}$ are three four-vectors given by
$$ n^{(i)\mu}(p)=\left\{ \frac{p^i}{M},\ \delta_{ij}+
\frac{p^ip^j}{M(E+M)}\right\}, \quad E=\sqrt{{\bf p}^2+M^2},
\qquad i,j=1,2,3,$$
$p_B=(E_B,{\bf Q})$ and $p_A=(E_A,{\bf P})$ are four-momenta of
initial and final mesons.

It is important to note that the wave functions entering the current
matrix element (\ref{mxet}) cannot  be both in the rest frame.
In the initial $B$ meson rest frame, the final $A$ meson is moving
with the recoil momentum ${\bf \Delta}$. The wave function
of the moving $A$ meson $\Psi_{A\,{\bf\Delta}}$ is connected
with the wave function in the rest frame
$\Psi_{A\,{\bf 0}}\equiv \Psi_{A}$ by the transformation \cite{f}
\begin{equation}
\label{wig}
\Psi_{A\,{\bf\Delta}}({\bf p})
=D_1^{1/2}(R_{L_{\bf\Delta}}^W)D_2^{1/2}(R_{L_{\bf\Delta}}^W)
\Psi_{A\,{\bf 0}}({\bf p}),
\end{equation}
where $R^W$ is the Wigner rotation, $L_{\bf\Delta}$ is the Lorentz boost
from the rest frame to a moving one, and the rotation matrix
$D^{1/2}(R)$ in the spinor representation  is given by 
\begin{equation}\label{d12}
{1 \ \ \,0\choose 0 \ \ \,1}D^{1/2}_{1,2}(R^W_{L_{\bf\Delta}})=
S^{-1}({\bf p}_{1,2})S({\bf\Delta})S({\bf p}),
\end{equation}
where
$$
S({\bf p})=\sqrt{\frac{\epsilon(p)+m}{2m}}\left(1+\frac{\bm{\alpha}
\cdot{\bf p}} {\epsilon(p)+m}\right)
$$
is the usual Lorentz transformation matrix of the four-spinor.

To calculate the radiative transition matrix element we adopt the
following procedure. We substitute the vertex functions $\Gamma^{(1)}$
and $\Gamma^{(2)}$  given by Eqs.~(\ref{gam1}) and (\ref{gam2})
in the decay matrix element (\ref{mxet})  and take into account the wave
function transformation (\ref{wig}).
The resulting structure of this matrix element is rather complicated,
because it is necessary  to integrate both over  $d^3 p$ and $d^3
q$. The $\delta$ function in expression  (\ref{gam1}) permits us to
perform one of these integrations and thus this contribution  can be easily
calculated. The calculation of the vertex function $\Gamma^{(2)}$
contribution is  more difficult. Here, instead
of a $\delta$ function, we have a complicated structure, containing the 
$Q\bar Q$ interaction potential in the meson. 
However, we can expand this contribution in powers of the heavy quark
velocities $v^2/c^2$ and then use the  quasipotential equation in
order to perform one of the integrations in the current matrix element.  
It is easy to see  that the vertex
function $\Gamma^{(2)}$ contributes already at  the first order of  
the $v^2/c^2$ expansion.

We consider two main types of radiative transitions: \\
a) Magnetic dipole
(M1) transitions which go with the spin flip of the quark ($\Delta
S=1$, $\Delta L=0$) and thus the initial and final states belong to
the same orbital excitation but have different
spins. Examples of such transitions are vector to pseudoscalar
($n\,{}^3\! S_1\to n'\,{}^1\! S_0 +\gamma$, $n\ge n'$) and
pseudoscalar to  vector
($n\,{}^1\! S_0\to n'\,{}^3\! S_1+\gamma$, $n > n'$) meson decays. \\
b)  Electric dipole (E1)
transitions in which the orbital quantum number is changed
($\Delta L=1$, $\Delta S=0$) and thus the initial and final states belong to
different orbital excitations but have the same
spin. Examples of such transitions are
$n\,{}^3\! S_1 \to n'\,{}^3\!  P_J\gamma$
($n>n'$) and $n\,{}^3\! P_J\to n'\,{}^3\! S_1\gamma$ ($n\ge n'$) decays.

\section{Radiative M1 transitions}

\label{m1}

\subsection{M1 decay rates}

The radiative M1 transition rate is given by \cite{gf}
\begin{equation}
  \label{eq:dr}
  \Gamma(B\to A+\gamma)=\frac{\omega^3}{3\pi}(2J'+1)
  \left|{\bm{\mathcal  M}}_{BA}\right|^2, 
  \quad {\rm where} \quad \omega=\frac{M_B^2-M_A^2}{2M_B},
\end{equation}
$M_B$ and $M_A$ are the initial and final meson masses, $J'$ is the
total angular momentum of the final meson. The matrix
element of the magnetic moment $\bm{\mathcal  M}$ is defined by
\begin{equation}
  \label{eq:magm} 
{\bm{\mathcal  M}}_{BA}=-\frac{i}2\left[\frac{\partial}{\partial{\bf
\Delta}}\times\left<A\left|{\bf
  J}(0)\right|B\right>\right]_{{\bf\Delta} =0}, 
\qquad {\bf \Delta}={\bf P}-{\bf Q}, 
\end{equation}
where $\left<A\left|J_\mu(0)\right|B\right>$ is the matrix element of the
electromagnetic current between initial ($B$) and final
($A$) meson states with  momenta ${\bf Q}$ and ${\bf P}$, respectively.

After inserting the vertex functions $\Gamma^{(1)}$  and $\Gamma^{(2)}$
from Eqs.~(\ref{gam1}) and (\ref{gam2})
in the decay matrix element (\ref{mxet}) with the account of the wave
function transformation (\ref{wig}), we carry out the expansion in
inverse powers of the heavy meson masses $M_{B,A}$, which are large due
to the presence of two heavy quarks $M_{B,A}\sim m_Q+m_{Q'}$. 
Then we calculate the matrix element of the magnetic
moment operator (\ref{eq:magm}) and get\\
(a) for the vector potential
\begin{eqnarray}
  \label{eq:muv}   
  {\bm{\mathcal  M}}_V& =& \int \frac{d^3 p}{(2\pi)^3}\bar \Psi_A({\bf
  p}) \frac{e_1}{2\epsilon_1(p)}\Biggl\{\bm{\sigma}_1+
  \frac{(1-\varepsilon)(1+2\kappa)[{\bf p}\times[\bm{\sigma}_1\times
  {\bf p}]]}{2\epsilon_1(p) 
  [\epsilon_1(p) +m_1]} \cr\nonumber\\
&&
+\frac{(1-\varepsilon)(1+\kappa)[{\bf p}\times[\bm{\sigma}_2\times
  {\bf p}]]}{\epsilon_1(p) 
  [\epsilon_2(p) +m_2]}\cr\nonumber\\
&&-\frac{\epsilon_2(p)}{M_B}\left(1+(1-\varepsilon)
  \frac{M_B-\epsilon_1(p)-\epsilon_2(p)}{\epsilon_1(p)}\right)i\left[{\bf
  p}\times \frac{\partial}{\partial{\bf p}}\right] \cr\nonumber\\
&& +\frac1{2M_B}\left[{\bf p}\times\left[{\bf p}\times\left(
  \frac{\bm{\sigma}_1}{\epsilon_1(p)+m_1}
  -\frac{\bm{\sigma}_2}{\epsilon_2(p)+m_2}\right)\right]\right]
\Biggr\}\Psi_B({\bf p}) +(1\leftrightarrow 2), 
\end{eqnarray}
(b) for the scalar potential
\begin{eqnarray}
  \label{eq:mus}   
{\bm{\mathcal  M}}_S& =& \int \frac{d^3 p}{(2\pi)^3}\bar \Psi_A({\bf
  p}) \frac{e_1}{2\epsilon_1(p)}\Biggl\{\left(1+
\varepsilon\;
\frac{\epsilon_1(p)+\epsilon_2(p)-M_B}{\epsilon_1(p)}\right)
\cr\nonumber\\
&&\times  \left(\bm{\sigma}_1- \frac{\epsilon_2(p)}{M_B}i\left[{\bf
  p}\times\frac{\partial}{\partial{\bf p}}\right] \right)
-  \frac{\varepsilon\;[{\bf p}\times[\bm{\sigma}_1\times {\bf p}]]}
  {2\epsilon_1(p) [\epsilon_1(p) +m_1]} \cr\nonumber\\
&& +\frac1{2M_B}\left[{\bf p}\times\left[{\bf p}\times\left(
  \frac{\bm{\sigma}_1} {\epsilon_1(p)+m_1}
  -\frac{\bm{\sigma}_2}{\epsilon_2(p)+m_2}\right)\right]\right]
\Biggr\}\Psi_B({\bf p}) +(1\leftrightarrow 2).
\end{eqnarray}
Note that the last terms in Eqs.~(\ref{eq:muv}), (\ref{eq:mus}) result
from the wave function transformation (\ref{wig}) from the moving
reference frame to the rest one. It is easy to see that in the
limit $p/m\to 0$ the usual nonrelativistic expression for the magnetic
moment follows.   

Since we are interested in radiative transitions between $S$ state
(vector and pseudoscalar) mesons, it is possible to evaluate
spin matrix elements
using the relation $\left<\bm{\sigma}_1\right>=-
\left<\bm{\sigma}_2\right>$. Then, taking into account that both quarks
are heavy ($Q$ and $Q'$), we further expand
Eqs.~(\ref{eq:muv}), (\ref{eq:mus}) in inverse powers of the heavy
quark mass $m_Q$ up to the second order corrections to the leading
contribution and get\\
(a) for the purely vector potential ($\varepsilon=0$) 
\begin{equation}
  \label{eq:muvex}   
{{\mathcal  M}}_V = \frac{e_Q}{2m_Q}\Biggl\{1-\frac{2\left<{\bf
  p}^2\right>}{3m_Q^2}+\frac{1+\kappa}3 \frac{\left<{\bf p}^2\right>}
{m_Q}\left(
  \frac1{m_Q} -\frac1{m_{Q'}}\right)
-\frac{\left<{\bf p}^2\right>}{6M_B}\left(\frac1{m_Q}+\frac1{m_{Q'}}
  \right)\Biggr\} -(Q\leftrightarrow Q'),
\end{equation}
(b) for the purely scalar potential ($\varepsilon=1$)
\begin{equation}
  \label{eq:musex}   
{{\mathcal  M}}_S =
\frac{e_Q}{2m_Q}\Biggl\{2-\frac{M_B-m_{Q'}}{m_Q}+
  \frac{\left<{\bf p}^2\right>}{2m_Q}\left(\frac1{m_{Q'}}-
  \frac1{3m_Q}\right)-\frac{\left<{\bf p}^2\right>}{6M_B}
\left(\frac1{m_Q}+\frac1{m_{Q'}} \right)\Biggr\} -(Q\leftrightarrow Q').
\end{equation}
Here $\left<\cdots\right>$ denotes the matrix element between radial
meson wave functions. For these matrix element calculations we use the
meson wave functions obtained calculating their mass spectra.

For the quarks of the same flavour ($m_Q=m_{Q'}$, $e_Q=-e_{\bar Q'}$)
and $\kappa=-1$ these expressions reduce further \cite{gf} \\
(a) for the purely vector potential ($\varepsilon=0$) 
\begin{equation}
  \label{eq:muvem}   
{{\mathcal  M}}_V = 2\frac{e_Q}{2m_Q}\Biggl\{1-\frac{2\left<{\bf
  p}^2\right>}{3m_Q^2}
-\frac{\left<{\bf p}^2\right>}{3M_Bm_Q}\Biggr\},
\end{equation}
(b) for the purely scalar potential ($\varepsilon=1$)
\begin{equation}
  \label{eq:musem}   
{{\mathcal  M}}_S =
2\frac{e_Q}{2m_Q}\Biggl\{3-\frac{M_B}{m_Q}+
  \frac{\left<{\bf p}^2\right>}{3m_Q}-\frac{\left<{\bf
        p}^2\right>}{3M_Bm_Q} \Biggr\}.
\end{equation}

\subsection{Results and discussion}

The resulting $M1$ radiative decay rates of charmonium, bottomonium
and $B_c$ are presented in Tables~\ref{tab:m1c}-\ref{tab:m1bc}. In
these Tables we give predictions both for allowed ($n\,{}^3\!S_1\to
n'\,{}^1\!S_0+\gamma$, $n'=n$) and hindered ($n>n'$)
decays. For the calculation of allowed decay rates we use expanded
expressions (\ref{eq:muvex})--(\ref{eq:musem}). For the hindered
transitions, which are strongly suppressed in the nonrelativistic
limit due to orthogonality of the initial and final state wave
functions, relativistic effects are decisive. Thus for their
calculations we use unexpanded expressions (\ref{eq:muv}) and
(\ref{eq:mus}). In Tables~\ref{tab:m1c}-\ref{tab:m1bc} we present the
photon energy $\omega$, 
the decay rates calculated discarding all relativistic corrections
$\Gamma^{\rm NR}$, as well as using relativistic expressions for
purely vector $\Gamma^V$, for purely scalar $\Gamma^S$ and for the
mixture (\ref{vlin}) of vector and scalar potentials $\Gamma$ with
$\varepsilon=-1$. Note that in all these calculations we use the
relativistic wave functions found calculating the meson mass spectra
in Sec.~\ref{sec:bc}.    

\begin{table}[htbp]
  \caption{Radiative M1 decay rates of charmonium. For decays
    involving $\eta_c'$ we give in parenthesis the results obtained
    using the recent value \cite{belle} of its mass. The values
    $\Gamma^{\rm exp}$ are taken from Ref.~\cite{pdg}. }
  \label{tab:m1c}
\begin{ruledtabular}
\begin{tabular}{ccccccc}
Decay& $\omega$ & $\Gamma^{\rm NR}$ & $\Gamma^V$ & $\Gamma^S$ & $\Gamma$ &
$\Gamma^{\rm exp} $ \\
     & MeV & keV & keV & keV & keV & keV \\
\hline
$J/\Psi \to \eta_c\gamma$ & 115 & 2.73 & 1.95 & 3.13 & 1.05 & $1.13\pm
0.35$\\ 
$\Psi'\to\eta_c'\gamma$ & 91(32) & 1.26(0.055) & 0.85(0.037) &
0.71(0.031) & 0.99(0.043) & 0.6 -- 3.9\footnote{This value from
  Ref.~\cite{cb} needs confirmation and is included only in the PDG
  Listings.} \\
$\Psi'\to \eta_c\gamma$ & 639 & 0.23 &0.61 & 0.35 & 0.95 & $0.84\pm
0.19$\\
$\eta_c'\to J/\Psi\gamma$& 463(514) & 0.26(0.36) &0.70(0.95)
&0.37(0.51) & 1.12(1.53) &
  \end{tabular}
\end{ruledtabular}  
\end{table}

\begin{table}[htbp]
  \caption{Radiative M1 decay rates of bottomonium.}
  \label{tab:m1b}
\begin{ruledtabular}
\begin{tabular}{ccccccc}
Decay& $\omega$ & $\Gamma^{\rm NR}$ & $\Gamma^V$ & $\Gamma^S$ &
     $\Gamma$ & $\cal B$  \\
     & MeV & eV & eV & eV & eV &($10^{-4}$)  \\
\hline
$\Upsilon\to\eta_b\gamma$ & 60 & 9.7 & 8.7 & 12.2 & 5.8& 1.1\\
$\Upsilon'\to\eta_b'\gamma$ & 33& 1.6 & 1.45 & 1.50 &1.40& 0.32\\
$\Upsilon''\to\eta_b''\gamma$ & 27& 0.9 & 0.8 & 0.8 &0.8& 0.30\\
$\Upsilon'\to\eta_b\gamma$ & 604& 1.3 & 3.4 & 1.3 & 6.4& 1.5\\
$\eta_b'\to\Upsilon\gamma$ & 516& 2.4 & 6.3 & 2.5 & 11.8&  \\
$\Upsilon''\to\eta_b\gamma$ & 911& 2.5 & 6.2 & 3.1 &10.5& 4.0\\
$\eta_b''\to\Upsilon\gamma$ & 831& 5.8 & 14.3 & 7.1 & 24.0&  \\
$\Upsilon''\to\eta_b'\gamma$ & 359& 0.2 & 0.6 & 0.1 &1.5& 0.57\\
$\eta_b''\to\Upsilon'\gamma$ & 301& 0.4 & 1.1 & 0.2 & 2.8& 
  \end{tabular}
\end{ruledtabular}  
\end{table}

\begin{table}[htbp]
  \caption{Radiative M1 decay rates of the $B_c$ meson.}
  \label{tab:m1bc}
\begin{ruledtabular}
\begin{tabular}{ccccccccc}
Transition& $\omega$ & $\Gamma^{\rm NR}$ & $\Gamma^V$ & $\Gamma^S$ &
     $\Gamma$ & $\Gamma$ \cite{eq}& $\Gamma$ \cite{gklt}& $\Gamma$
     \cite{ful}\\ 
     & MeV & eV & eV & eV & eV & eV & eV & eV\\
\hline
$1\,{}^3\! S_1\to 1\,{}^1\! S_0\gamma$ & 62 & 73 & 48 & 66 & 33
& 135 & 60 &59\\
$2\,{}^3\! S_1\to 2\,{}^1\! S_0\gamma$ & 46& 30 & 24 & 32 &17 & 29 &
     10 & 12\\ 
$2\,{}^3\! S_1\to 1\,{}^1\! S_0\gamma$ & 584& 141 & 412 & 398 & 428 &
     123 & 98 & 122\\ 
$2\,{}^1\! S_0\to 2\,{}^3\! S_1\gamma$ & 484& 160 & 471 & 454 & 488 &
     93 & 96 & 139 
  \end{tabular}
\end{ruledtabular}  
\end{table}

The M1 radiative decay rates are very
sensitive to relativistic effects. Even for allowed transitions
relativistic and nonrelativistic results differ significantly. An
important example is the decay $J/\Psi \to \eta_c\gamma$. It
is well known that the nonrelativistic predictions for its rate are
more than two times larger than the experimental data. As we see
from Table~\ref{tab:m1c}, the inclusion of the relativistic corrections
for purely scalar or purely vector potentials do not bring theoretical
results in agreement with experiment. For the purely scalar potential
the decay rate even increases by 15\%. On the other hand for the purely
vector potential relativistic effects decrease the decay rate by 25\%,
but such decrease is not enough: the theoretical result still deviates
from experimental data by more than $2\sigma$.~\footnote{This is
compatible with the estimate that relativistic effects can give
contributions of order of 20--30\% in charmonium.} Only for the
mixture of vector and scalar potentials (\ref{vlin}) we get the
necessary decrease of the decay rate which brings theory in 
agreement with experimental data for the $J/\Psi \to \eta_c\gamma$
decay rate. For the hindered decay $\Psi'\to \eta_c\gamma$ the decay
rate calculated for the mixture of vector and scalar potentials is
also in good agreement with experiment while the rates for pure
potentials (especially the scalar one) are lower than the experimental
value. In Table~\ref{tab:m1c} we give predictions for decays involving
the first radial excitation of the pseudoscalar state
$\eta_c'(2\,{}^1\!S_0)$ as well. Since there
are two contradicting experimental measurements of its mass we
calculated the rates using both values. The results obtained using
recent Belle value \cite{belle} are given in parenthesis.

In Table~\ref{tab:m1b} predictions for M1 decay rates and branching
fractions of bottomonium are given. Since the hyperfine splitting in
bottomonium is 
predicted to be small (around 60 MeV, see Table~\ref{bottom}) the
photon energies and hence decay rates of allowed M1 transitions are
very small. This is one of the main reasons why no spin-singlet
$S$-wave levels $\eta_b(n\,{}^1\!S_0)$ have been observed yet. 
Our results show that relativistic effects for the favored mixture of
vector and scalar potentials further decrease the allowed M1 decay rates. 

Recently it was argued by Godfrey and Rosner \cite{grnb} that
hindered transitions could be more favorable for discovering the
$\eta_b$. Their analysis of different quark model predictions showed
that most nonrelativistic models favor the $\eta_b(1S)$ production from
$\Upsilon(2S)$ decays, while the account of relativistic corrections
makes prospects for discovering $\eta_b(1S)$ in $\Upsilon(3S)$
radiative decays comparable to those in $\Upsilon(2S)$ decays. Our
present relativistic consideration of these decays supports this
observation. Indeed we see from Table~\ref{tab:m1b} that 
relativistic effects significantly increase rates of hindered
transitions. The hindered decay
$\Upsilon''(3S)\to \eta_b(1S)\gamma$ has the largest branching
fraction $4.0\times 10^{-4}$, which is almost 2.7 times larger than
the $\Upsilon'(2S)\to \eta_b(1S)\gamma$ decay branching fraction. Very
recently CLEO Collaboration \cite{cleonb} searched for $\eta_b(1S)$ in
such hindered M1 transition from $\Upsilon''(3S)$. No evidence of
$\Upsilon''(3S)\to \eta_b(1S)\gamma$ transitions was found and rather
strict upper limits on the branching fraction were set: ${\cal
  B}(\Upsilon''(3S)\to \eta_b(1S)\gamma) < 6 \times 10^{-4}$, which
rule out many previous phenomenological predictions reviewed in
Ref.~\cite{grnb}. Our model result for the branching fraction of
this decay is below but rather close to this experimental upper
limit. 

In Table~\ref{tab:m1bc} we give predictions for decay rates of M1
radiative transitions of the $B_c$ meson in our model in comparison with
previous nonrelativistic quark model analysis \cite{eq,gklt,ful}. We
see that relativistic effects play an important role in $B_c$ meson M1
radiative decays. They reduce the rates of allowed decays and increase
the rates of hindered transitions. The largest rates are predicted for
the latter decays which are increased by relativistic effects almost
by the factor of 3 and thus they are an order of magnitude larger than
the rates of allowed M1 transitions.       

\section{Radiative E1 transitions}

\label{e1}

\subsection{E1 decay rates}

The radiative E1 transition rate is given by \cite{gmf}
\begin{equation}
  \label{eq:e1r}
  \Gamma(B\to A+\gamma)=\frac{\omega^3}{3\pi} |{\bf D}_{BA}|^2,
 \quad {\rm where} \quad \omega=\frac{M_B^2-M_A^2}{2M_B},
\end{equation}
$M_B$ and $M_A$ are the initial and final meson masses.
The matrix
element of the electric dipole moment ${\bf D}_{BA}$ is defined by
\begin{equation}
  \label{eq:egm}
  {{\bf D}_{BA}}=-i\left. \frac{\partial}{\partial{\bf
  \Delta}}\left<A\left|{J_0}(0)\right|B\right>\right|_{\bf\Delta =0},
\qquad {\bf \Delta}={\bf P}-{\bf Q}, 
\end{equation}
where $\left<A\left|J_\mu(0)\right|B\right>$ is the matrix element of the
electromagnetic current between initial ($B$) and final
($A$) meson states with  momenta ${\bf Q}$ and ${\bf P}$, respectively.

We substitute expressions (\ref{mxet})--(\ref{gam2}) in the definition
of the electric dipole moment (\ref{eq:egm}) and take into account
the relativistic transformation of the wave function (\ref{wig}). Then,
discarding some terms of order $v^4/c^4$ and higher, we get the following
expressions for the electric dipole moment ${\bf D}_{BA}$ \cite{gmf}
(indices 1,2 are changed to $Q,Q'$)

\noindent (a) for the purely vector potential
\begin{eqnarray}
  \label{eq:dv}
  {\bf D}_{V}& =& \int \frac{d^3 p}{(2\pi)^3}\bar \Psi_A({\bf
  p})e_Q\Biggl\{i\,\frac{\epsilon_{Q'}(p)}{M_B}\frac{\partial}{\partial{\bf
  p}} -\frac{\left[\bm{\sigma}_{Q}\times
  {\bf p}\right]}{2\epsilon_{Q}(p)[\epsilon_{Q}(p)+m_Q]}
  \Biggl(1-\frac{\epsilon_{Q}(p)}{M_B}\cr\nonumber\\
&&+\frac{2[M_B-\epsilon_Q(p)-\epsilon_{Q'}(p)]}{m_Q}\Biggr)
-\frac{\left[\bm{\sigma}_{Q'}\times
  {\bf p}\right]}{2M_B[\epsilon_{Q'}(p)+m_{Q'}]}\Biggr\}\Psi_B({\bf p}) -
(Q\leftrightarrow Q'),
\end{eqnarray}
(b) for the purely scalar potential
\begin{eqnarray}
  \label{eq:ds}
  {\bf D}_{S}& =& \int \frac{d^3 p}{(2\pi)^3}\bar \Psi_A({\bf
  p})e_Q\Biggl\{i\,\frac{\epsilon_{Q'}(p)}{M_B}\frac{\partial}{\partial{\bf
  p}} -\frac{\left[\bm{\sigma}_{Q}\times
  {\bf p}\right]}{2\epsilon_{Q}(p)[\epsilon_{Q}(p)+m_Q]}
  \Biggl(1-\frac{\epsilon_{Q}(p)}{M_B}\cr\nonumber\\
&&-\frac{2[M_B-\epsilon_Q(p)-\epsilon_{Q'}(p)]}{m_Q}\Biggr)
-\frac{\left[\bm{\sigma}_{Q'}\times
  {\bf p}\right]}{2M_B[\epsilon_{Q'}(p)+m_{Q'}]}\Biggr\}\Psi_B({\bf p}) -
(Q\leftrightarrow Q').
\end{eqnarray}
The operator $i{\partial}/{\partial{\bf p}}$ in Eqs.~(\ref{eq:dv}),
(\ref{eq:ds}) corresponds in the coordinate space to the operator
${\bf r}$. All other terms in these equations are relativistic corrections.
Thus in the nonrelativistic limit the standard expression for the
electric dipole moment is recovered.   

It is easy to see that there are three different structures with
respect to the orbital variables in Eqs.~(\ref{eq:dv}), (\ref{eq:ds}):
${\bf r}$, $[(\bm{\sigma_Q}+\bm{\sigma}_{Q'})\times {\bf p}]$ and
$[(\bm{\sigma_Q}-\bm{\sigma}_{Q'})\times {\bf p}]$. Thus the matrix
element of the electric dipole moment  for the electromagnetic
transition $nJMLS \to n'J'M'L'S'+\gamma$ can be presented in the form
\begin{equation}
  \label{eq:dme}
  {\bf D}_{V,S}=\langle n'  J' M' L' S'|
  {\cal A}(p^2){\bf r}-{\cal B}^{V,S}(p^2)[{\bf S}\times {\bf
  p}]-{\cal C}^{V,S}(p^2)[({\bf S}_Q-{\bf
  S}_{Q'}) \times {\bf p}]|nJMLS\rangle, 
\end{equation}
where functions ${\cal A}(p^2)$, ${\cal B}^{V,S}(p^2)$ and ${\cal
  C}^{V,S}(p^2)$ up to 
order ${\bf p}^2/m_{Q(Q')}^2$ are given by
\begin{eqnarray}
{\cal A}(p^2)&=&\frac{e_Qm_{Q'}-e_{Q'}m_{Q}}{M_B}+ 
\frac{e_Qm_Q-e_{Q'}m_{Q'}}{M_B}\frac{{\bf p}^2}{2m_Qm_{Q'}},\\ \nonumber\\
{\cal B}^V(p^2)&=&\frac{e_Q}{4m_Q^2}\left(1+\frac{2(M-m_Q-m_{Q'})}{m_Q}
  -\frac{7{\bf p}^2}{4m_Q^2}-\frac{{\bf p}^2}{m_{Q'}^2}\right)\cr\nonumber\\
&&-\frac{e_{Q'}}{4m_{Q'}^2}\left(1+\frac{2(M-m_Q-m_{Q'})}{m_{Q'}}
  -\frac{7{\bf p}^2}{4m_{Q'}^2}-\frac{{\bf p}^2}{m_Q^2}\right)\cr\nonumber\\
&&-\frac{e_Q+e_{Q'}}{4M_B}\left[\frac1{m_Q}-\frac1{m_{Q'}}-\frac{{\bf
    p}^2}4 \left(\frac1{m_Q^3}-\frac1{m_{Q'}^3}\right)\right],\\\nonumber\\
{\cal C}^V(p^2)&=&\frac{e_Q}{4m_Q^2}\left(1+\frac{2(M-m_Q-m_{Q'})}{m_Q}
  -\frac{7{\bf p}^2}{4m_Q^2}-\frac{{\bf p}^2}{m_{Q'}^2}\right)\cr\nonumber\\
&&+\frac{e_{Q'}}{4m_{Q'}^2}\left(1+\frac{2(M-m_Q-m_{Q'})}{m_{Q'}}
  -\frac{7{\bf p}^2}{4m_{Q'}^2}-\frac{{\bf p}^2}{m_Q^2}\right)\cr\nonumber\\
&&-\frac{e_Q+e_{Q'}}{4M_B}\left[\frac1{m_Q}+\frac1{m_{Q'}}-\frac{{\bf
    p}^2}4 \left(\frac1{m_Q^3}+\frac1{m_{Q'}^3}\right)\right],\\\nonumber\\
{\cal B}^S(p^2)&=&\frac{e_Q}{4m_Q^2}\left(1-\frac{2(M-m_Q-m_{Q'})}{m_Q}
  +\frac{{\bf p}^2}{4m_Q^2}+\frac{{\bf p}^2}{m_Qm_{Q'}}\right)\cr\nonumber\\
&&-\frac{e_{Q'}}{4m_{Q'}^2}\left(1-\frac{2(M-m_Q-m_{Q'})}{m_{Q'}}
  +\frac{{\bf p}^2}{4m_{Q'}^2}+\frac{{\bf p}^2}{m_Qm_{Q'}}\right)\cr\nonumber\\
&&-\frac{e_Q+e_{Q'}}{4M_B}\left[\frac1{m_Q}-\frac1{m_{Q'}}-\frac{{\bf
    p}^2}4 \left(\frac1{m_Q^3}-\frac1{m_{Q'}^3}\right)\right],\\\nonumber\\
{\cal C}^S(p^2)&=&\frac{e_Q}{4m_Q^2}\left(1-\frac{2(M-m_Q-m_{Q'})}{m_Q}
  +\frac{{\bf p}^2}{4m_Q^2}+\frac{{\bf p}^2}{m_Qm_{Q'}}\right)\cr\nonumber\\
&&+\frac{e_{Q'}}{4m_{Q'}^2}\left(1-\frac{2(M-m_Q-m_{Q'})}{m_{Q'}}
  +\frac{{\bf p}^2}{4m_{Q'}^2}+\frac{{\bf p}^2}{m_Qm_{Q'}}\right)\cr\nonumber\\
&&-\frac{e_Q+e_{Q'}}{4M_B}\left[\frac1{m_Q}+\frac1{m_{Q'}}-\frac{{\bf
    p}^2}4 \left(\frac1{m_Q^3}+\frac1{m_{Q'}^3}\right)\right].
\end{eqnarray}
The last structure in Eq.~(\ref{eq:dme}) proportional to $[({\bf S}_Q-{\bf
  S}_{Q'}) \times {\bf p}]$ leads to the spin-flip transitions.
It vanishes for the $c\bar c$ and $b\bar b$ mesons,
consisting of the quark and antiquark of the same flavour
($m_Q=m_{Q'}$, $e_Q=-e_{Q'}$) since in that case ${\cal
  C}^{V,S}(p^2)=0$. The functions ${\cal A}(p^2)$, ${\cal
  B}^{V,S}(p^2)$ then simplify and coincide with the ones found
previously \cite{gmf}
\begin{eqnarray*}
  {\cal A}(p^2)&=&e_Q\left(\frac{2m_Q}{M_B} +\frac{{\bf p}^2}{M_Bm_Q}
\right),\cr\nonumber\\
{\cal B}^V(p^2)&=&\frac{e_Q}{2m_Q^2}\left(1+\frac{2(M_B-2m_Q)}{m_Q} -
  \frac{11{\bf p}^2}{4m_Q^2} \right),\cr\nonumber\\
{\cal B}^S(p^2)&=&\frac{e_Q}{2m_Q^2}\left(1-\frac{2(M_B-2m_Q)}{m_Q} +
  \frac{5{\bf p}^2}{4m_Q^2} \right).
\end{eqnarray*}
In the case of the $B_c$ meson ${\cal C}^{V,S}(p^2)\ne 0$ and thus relativistic
corrections lead to spin-flip transitions ($S'=S\pm1$) but only for decays
involving mixed states $nP1$, $nP1'$ (\ref{eq:pmix}) or $nD2$, $nD2'$
(\ref{eq:dmix}). For all other transitions the spin-flip correction
vanishes due to momenta relations (see Eq.~(\ref{eq:chi}) below).   

Using the Wigner-Eckart theorem and relations for matrix
elements of the tensor operator between coupled functions, one can
rewrite Eq.~(\ref{eq:dme}) in the form 
\begin{eqnarray}
  \label{eq:adme}
{D}_i&=& (-1)^{J'+J+L+S-M'}\left(
  \begin{array}{ccc}
J' & 1 & J \\
-M' & i & M
  \end{array}
\right) \sqrt{(2J+1)(2J'+1)}\cr\nonumber\\
&&
\times\Bigglb[\left\{
      \begin{array}{ccc}
L' & J' & S \\
J & L & 1
      \end{array}
\right\} \delta_{SS'} 
\left(\left<n'L'||{\cal A}(p^2) r||n L\right>-\eta(J',L',J,L,S)
  \left<n'L'||{\cal B}(p^2)p||n L\right> \right)\cr\nonumber\\
&&\qquad-\chi(J',L',J,L) \delta_{S S'\pm1}\left<n'L'||{\cal C}(p^2)p
    ||n L\right> \Biggrb],
\end{eqnarray}
where
\begin{equation}
  \label{eq:eta}
  \eta(J',L',J,L,S)=(-1)^{L+J+1}\sqrt{6}[S(S+1)(2S+1)]^{1/2}\left(
  \left\{ 
      \begin{array}{ccc}
L' & J' & S \\
J & L & 1
      \end{array}
\right\}\right)^{-1}
\left\{
    \begin{array}{cccc}
J & S & L \\
J' & S & L' \\
1 & 1 & 1 
    \end{array}
\right\},
\end{equation}
\begin{equation}
  \label{eq:chi}
\chi(J',L',J,L)=(-1)^{L+L'}\sqrt{\frac{2}{2J'+1}} \left\{ 
      \begin{array}{ccc}
J & L & 1 \\
1 & 1 & L'
      \end{array}
\right\}. 
\end{equation}
Here 
$$ \left(
  \begin{array}{ccc}
J' & 1 & J \\
-M' & q & M
  \end{array}
\right), \qquad \left\{
      \begin{array}{ccc}
L' & J' & S \\
J & L & 1
      \end{array}
\right\}, \quad {\rm and} \quad \left\{
    \begin{array}{cccc}
J & S & L \\
J' & S & L' \\
1 & 1 & 1 
    \end{array}
\right\}$$
are $3j$-, $6j$- and $9j$-symbols, $\left<n'L'||\cdots||n L\right>$
are reduced matrix elements. 

The total E1 decay rate of the $nJLS$ state is obtained by summing the
decay rates (\ref{eq:e1r}) over all possible values of $M'$ for a
fixed value of $M$. The resulting expression is
\begin{eqnarray}
  \label{eq:edr}
  \Gamma^{V,S}(nJLS\to n'J'L'S'+\gamma)=\frac43\alpha\,
  \omega^3&\!\!\!\Biggl\{&\!\!\!\delta_{SS'}C^{1/2}(J',L',J,L,S)
\Biggl[(L'-L)\left<R_{n'L'}
  \left|{\cal A}(p^2)r\right|R_{nL}\right>\cr\nonumber\\
&&
-\frac{\eta(J',L',J,L,S)}{\sqrt{\max(L',L)}} \left<n'L'||
{\cal B}^{V,S}(p^2)p||nL\right> \Biggr]\cr\nonumber\\
&& - \delta_{SS'\pm1}\chi(J',L',J,L)\left<n'L'||{\cal
  C}^{V,S}(p^2)p||nL\right> \Biggl\}^2,
\end{eqnarray}
where
$$C(J',L',J,L,S)=\max(L',L)(2J'+1)\left\{ 
      \begin{array}{ccc}
L' & J' & S \\
J & L & 1
      \end{array}
\right\}^2.$$
The reduced matrix elements can be expressed through the usual matrix
elements over radial wave functions $R_{nL}(r)$: \\
(a) for the transitions between $P$ and $S$ states
\begin{eqnarray}
  \label{eq:psrm}
  \left<n'0||p||n1\right>&=&-\left<R'_{n'S}|R_{nP}\right>,\cr\nonumber\\
\left<n'0||{\bf p}^2p||n1\right>&=&\left<R'''_{n'S}|R_{nP}\right>+
2\left<R''_{n'S}\left|\frac1{r}\right|R_{nP}\right>
-2\left<R'_{n'S}\left|\frac1{r^2}\right|R_{nP}\right>;
\end{eqnarray}
(b) for the transitions between $D$ and $P$ states
\begin{eqnarray}
  \label{eq:dprm}
  \left<n'1||p||n2\right>&=&-\sqrt{2}\left(\left<R_{n'P}\left|
 \frac1{r}\right|R_{nD}\right>-\left<R'_{n'P}|R_{nD}\right>\right),
\cr\nonumber\\ 
\left<n'1||{\bf p}^2p||n2\right>&=&-\sqrt{2}\Biggl(
\left<R'''_{n'P}|R_{nD}\right>+
\left<R''_{n'P}\left|\frac1{r}\right|R_{nD}\right>\cr\nonumber\\
&&
-6\left<R'_{n'P}\left|\frac1{r^2}\right|R_{nD}\right>
+6\left<R_{n'P}\left|\frac1{r^3}\right|R_{nD}\right>\Biggr),
\end{eqnarray}
where the prime means differentiation of $R$ with respect to $r$.

\subsection{Results and discussion}

\begin{table}
  \caption{Radiative E1 transition rates of charmonium.  For decays
    involving $\eta_c'$ we give in parenthesis the results obtained
    using the recent value \cite{belle} of its mass.}
  \label{tab:e1c}
\begin{ruledtabular}
\begin{tabular}{ccccccc}
Decay& $\omega$ & $\Gamma^{\rm NR}$ & $\Gamma^V$ & $\Gamma^S$ & $\Gamma$ &
     $\Gamma^{\rm exp}$ \cite{pdg} \\
     & MeV & keV & keV & keV & keV & keV \\
\hline
$2\,{}^3\! S_1\to1\,{}^3\! P_0\gamma$& 259 & 51.7 &34.6 &44.0 & 26.3 &
     $26.1\pm 3.2$\\
$2\,{}^3\! S_1\to1\,{}^3\! P_1\gamma$& 171 & 44.9 & 30.1 & 38.3 & 22.9
     & $25.2\pm 3.0$\\
$2\,{}^3\! S_1\to1\,{}^3\! P_2\gamma$& 128 & 30.9 & 22.9 & 28.1 & 18.2
     & $20.4 \pm 2.5$\\
$2\,{}^1\! S_0\to1\,{}^1\! P_1\gamma$& 68(128) & 8.6(57)& 6.2(41) &
     6.2(41) & 6.2(41)&\\
$1\,{}^3\! P_0\to1\,{}^3\! S_1\gamma$& 305 & 161 & 151 & 184 & 121 &
     $165\pm36$\\ 
$1\,{}^3\! P_1\to1\,{}^3\! S_1\gamma$& 389 & 333 & 285 & 305 & 265 &
     $291\pm 51$\\ 
$1\,{}^3\! P_2\to1\,{}^3\! S_1\gamma$& 430 & 448 & 309 & 292 & 327 &
     $389\pm 52$\\ 
$1\,{}^1\! P_1\to1\,{}^1\! S_0\gamma$& 504 & 723 & 560 & 560 & 560 &  \\
$1\,{}^3\! D_1\to1\,{}^3\! P_0\gamma$& 361 & 423 & 344 & 334 & 355 &  \\
$1\,{}^3\! D_1\to1\,{}^3\! P_1\gamma$& 277 & 142 & 127 & 120 & 135 &  \\
$1\,{}^3\! D_1\to1\,{}^3\! P_2\gamma$& 234 & 5.8 & 6.2 & 5.6 & 6.9 &  \\
$1\,{}^3\! D_2\to1\,{}^3\! P_1\gamma$& 291 & 297 & 215 & 215 & 215 &  \\
$1\,{}^3\! D_2\to1\,{}^3\! P_2\gamma$& 248 & 62  & 55  & 51  & 59  &  \\
$1\,{}^3\! D_3\to1\,{}^3\! P_2\gamma$& 250 & 252 & 163 & 170 & 156 &  \\
$1\,{}^1\! D_2\to1\,{}^1\! P_1\gamma$& 275 & 335 & 245 & 245 & 245 &  
\end{tabular}
\end{ruledtabular}
\end{table}

The results of numerical calculations of charmonium E1 radiative decay
rates using Eqs.~(\ref{eq:eta})--(\ref{eq:dprm}) are presented in
Table~\ref{tab:e1c}. For calculations of photon energies $\omega$ we
used the experimentally measured masses of charmonium $S$ and $P$
states.~\footnote{For decays involving $\eta_c'(2S)$, as in the case
of M1 transitions, we use both experimental values of its  
mass, giving a
prediction for the recent Belle value \cite{belle} in parenthesis.}  
For masses of $D$ states we used our model predictions from
Table~\ref{charm}. We give predictions for decay rates calculated in
the nonrelativistic limit $\Gamma^{\rm NR}$, for relativistic decay
rates with pure vector $\Gamma^V$ and scalar $\Gamma^S$ potentials as
well as for the mixture (\ref{vlin}) of vector and scalar potentials
$\Gamma$ with $\varepsilon=-1$. As in the case of M1 decay rates
calculations, we use the relativistic wave functions in our numerical
analysis.

The results presented in Table~\ref{tab:e1c} show that relativistic
effects play an important role in E1 decays of charmonium. The most
sensitive to the relativistic corrections are decays $\Psi'(2S)\to
\chi_{cJ}(1P)+\gamma$. Their account leads to the considerable reduction
of the  decay rates. The rates for the vector potential are reduced
more significantly than for the
scalar one.  As a result, there arises an approximately twofold reduction
of decay rates for the mixture of vector and scalar potentials with
the value of mixing parameter $\varepsilon=-1$, bringing theoretical
predictions in good agreement with experimental data. The large
influence of relativistic corrections originates from the fact that the
zero of the $2S$ wave function is close to the maximum of the $1P$ wave
function. This results in a reduction of the leading order decay matrix
element $\left<1P|r|2S|\right>$.  Therefore relatively small
relativistic corrections produce such a large effect. This observation
is confirmed by the calculations of the $\chi_{cJ}(1P)\to
J/\Psi+\gamma$ decay rates. Here both initial and final states do not
have zeros and 
relativistic contributions have usual values and lead to an approximately
$25\%$ reduction of the decay rate. All theoretical predictions are in
nice agreement with data. In Table~\ref{tab:e1c} we also give
predictions for E1 decay rates of charmonium $1D$ states. At present
only the $1\,{}^3\! D_1$ state is experimentally observed. This state is
considerably broader, since it lies above the $D\bar D$ threshold. The
observed state $\Psi(3770)$ can also have a significant $2\,{}^3\!
S_1$ state admixture \cite{r}. If we consider it to be a pure $D$
state, then using its measured total decay rate, we get the following
predictions for the E1 radiative decay branching fractions:
\[
{\cal B}(1\,{}^3\! D_1\to1\,{}^3\! P_0\gamma)\approx 1.5 \%; \quad
{\cal B}(1\,{}^3\! D_1\to1\,{}^3\! P_1\gamma)\approx 0.6 \%; \quad
{\cal B}(1\,{}^3\! D_1\to1\,{}^3\! P_2\gamma)\approx 0.03 \%.
\]
On the other hand, the tensor $1\,{}^3\! D_2$ and $1\,{}^1\! D_2$
charmonium states are under the threshold of open charm production, since
their masses are slightly below the $D\bar D^*$ threshold, and the
decay of these 
states to $D\bar D$ is forbidden by parity and angular momentum
conservation. Thus E1 radiative transitions are the main decay
channels of these states.

\begin{table}
  \caption{Radiative E1 transition rates of bottomonium.}
  \label{tab:e1b}
\begin{ruledtabular}
\begin{tabular}{ccccccc}
Decay& $\omega$ & $\Gamma^{\rm NR}$ & $\Gamma^V$ & $\Gamma^S$ & $\Gamma$ &
     $\Gamma^{\rm exp}$ \cite{pdg} \\
     & MeV & keV & keV & keV & keV & keV \\
\hline
$2\,{}^3\! S_1\to1\,{}^3\! P_0\gamma$& 162 & 1.65 & 1.64 & 1.66 & 1.62
    & $1.67\pm 0.37$\\
$2\,{}^3\! S_1\to1\,{}^3\! P_1\gamma$& 130 & 2.57 & 2.48 & 2.51 & 2.45
     & $2.9\pm 0.6$\\
$2\,{}^3\! S_1\to1\,{}^3\! P_2\gamma$& 109 & 2.53 & 2.49 & 2.52 & 2.46
     & $3.08\pm 0.6$\\
$2\,{}^1\! S_0\to1\,{}^1\! P_1\gamma$& 98 & 3.25 & 3.09 & 3.09 & 3.09 &   \\
$1\,{}^3\! P_0\to1\,{}^3\! S_1\gamma$& 391 & 29.5 & 30.6 & 31.3 & 29.9 &  \\
$1\,{}^3\! P_1\to1\,{}^3\! S_1\gamma$& 422 & 37.1 & 37.0 & 37.4 & 36.6 &  \\
$1\,{}^3\! P_2\to1\,{}^3\! S_1\gamma$& 442 & 42.7 & 39.8 & 39.3 & 40.2 &  \\
$1\,{}^1\! P_1\to1\,{}^1\! S_0\gamma$& 480 & 54.4 & 52.6 & 52.6 & 52.6 &  \\
$3\,{}^3\! S_1\to2\,{}^3\! P_0\gamma$& 123 & 1.65 & 1.51 & 1.52 & 1.49
     & $1.42\pm    0.25$\\
$3\,{}^3\! S_1\to2\,{}^3\! P_1\gamma$& 100 & 2.65 & 2.43 & 2.45 & 2.41
     & $2.97\pm   0.43$\\
$3\,{}^3\! S_1\to2\,{}^3\! P_2\gamma$& 86  & 2.89 & 2.69 & 2.71 & 2.67
     & $3.0\pm 0.45$\\
$3\,{}^1\! S_0\to2\,{}^1\! P_1\gamma$& 73 & 3.07 & 2.78 & 2.78 & 2.78 &  \\
$3\,{}^3\! S_1\to1\,{}^3\! P_0\gamma$& 484 & 0.124 & 0.040 & 0.054 &0.027 & \\
$3\,{}^3\! S_1\to1\,{}^3\! P_1\gamma$& 453 & 0.307 & 0.097 & 0.134 & 0.067 &
     $0.041\pm 0.029$\footnotemark\footnotetext{Needs confirmation.}\\
$3\,{}^3S_1\to1\,{}^3P_2\gamma$& 433 & 0.445 & 0.141 & 0.195 & 0.097 &
     $0.064\pm 0.045$\footnotemark[1]\\
$3\,{}^1\! S_0\to1\,{}^1\! P_1\gamma$& 427 & 0.770 & 0.348 & 0.348 & 0.348 & \\
$2\,{}^3\! P_0\to2\,{}^3\! S_1\gamma$& 207 & 11.7 & 11.1 & 11.2 & 11.0 &  \\
$2\,{}^3\! P_1\to2\,{}^3\! S_1\gamma$& 230 & 15.9 & 14.8 & 14.8 & 14.7 & \\
$2\,{}^3\! P_2\to2\,{}^3\! S_1\gamma$& 243 & 18.8 & 16.7 & 16.6 & 16.7 & \\
$2\,{}^1\! P_1\to2\,{}^1\! S_0\gamma$& 262 & 23.6 & 21.4 & 21.4 & 21.4 & \\
$2\,{}^3\! P_0\to1\,{}^3\! S_1\gamma$& 743 & 7.36 & 7.58 & 8.41 & 6.79 & \\
$2\,{}^3\! P_1\to1\,{}^3\! S_1\gamma$& 764 & 8.01 & 7.90 & 8.33 & 7.49 & \\
$2\,{}^3\! P_2\to1\,{}^3\! S_1\gamma$& 776 & 8.41 & 7.61 & 7.20 & 8.02 & \\
$2\,{}^1\! P_1\to1\,{}^1\! S_0\gamma$& 820 & 9.9 & 9.36 & 9.36 & 9.36 & 
\end{tabular}
\end{ruledtabular}
\end{table}

\begin{table}[htb]
  \caption{Radiative E1 transition rates of bottomonium involving $D$
     states.} 
  \label{tab:e1b2}
\begin{ruledtabular}
\begin{tabular}{cccccc}
Decay& $\omega$ & $\Gamma^{\rm NR}$ & $\Gamma^V$ & $\Gamma^S$ & $\Gamma$ \\
     & MeV & keV & keV & keV & keV \\
\hline
$1\,{}^3\! D_1\to1\,{}^3\! P_0\gamma$& 280 & 24.2 & 23.4 & 23.4 & 23.4  \\
$1\,{}^3\! D_1\to1\,{}^3\! P_1\gamma$& 256 & 12.9 & 12.7 & 12.7 & 12.7  \\
$1\,{}^3\! D_1\to1\,{}^3\! P_2\gamma$& 235 & 0.67 & 0.69 & 0.69 & 0.69  \\
$1\,{}^3\! D_2\to1\,{}^3\! P_1\gamma$& 262 & 24.8 & 23.3 & 23.3 & 23.3  \\
$1\,{}^3\! D_2\to1\,{}^3\! P_2\gamma$& 241 & 6.45 & 6.35 & 6.35 & 6.35  \\
$1\,{}^3\! D_3\to1\,{}^3\! P_2\gamma$& 244 & 26.7 & 24.6 & 24.6 & 24.6  \\
$1\,{}^1\! D_2\to1\,{}^1\! P_1\gamma$& 254 & 30.2 & 28.4 & 28.4 & 28.4  \\
$2\,{}^3\! P_0\to1\,{}^3\! D_1\gamma$& 81 & 1.17 & 1.16 & 1.16 & 1.17  \\
$2\,{}^3\! P_1\to1\,{}^3\! D_1\gamma$& 104 & 0.62 & 0.61 & 0.60 & 0.615 \\
$2\,{}^3\! P_1\to1\,{}^3\! D_2\gamma$& 98 & 1.56 & 1.56 & 1.56 & 1.56  \\
$2\,{}^3\! P_2\to1\,{}^3\! D_1\gamma$& 117 & 0.036 & 0.034 & 0.034 & 0.035  \\
$2\,{}^3\! P_2\to1\,{}^3\! D_2\gamma$& 111 & 0.454 & 0.446 & 0.443 & 0.449  \\
$2\,{}^3\! P_2\to1\,{}^3\! D_3\gamma$& 108 & 2.34 & 2.36 & 2.37 & 2.35  \\
$2\,{}^1\! P_1\to1\,{}^1\! D_2\gamma$& 102 & 2.62 & 2.43 & 2.43 & 2.43  
\end{tabular}
\end{ruledtabular}
\end{table}

The calculated decay rates of  E1 radiative transitions in bottomonium
are presented in Tables~\ref{tab:e1b} and \ref{tab:e1b2}. The
influence of relativistic effects in bottomonium is considerably less
than in charmonium. The contribution of relativistic corrections does not
exceed 10\% almost for all decays. The only exceptions are decays
$\Upsilon''(3S)\to \chi_{bJ}(1P)+\gamma$, where the leading
contribution is substantially reduced due to the significantly
different number of zeros in initial $3S$ and final $1P$ wave
functions. For all $S\to P+\gamma$ transitions we find good agreement
of our model predictions with experimental data.  

The comparison of the theoretical predictions for the radiative decays
of $P$ states of bottomonium $\chi_b(nP)$ with the experimental data is
complicated by the fact that the total decay rates of these states are not
measured yet. Experiment gives only branching fractions ${\cal B}\equiv
\Gamma[\chi_b(nP)\to \Upsilon(n'S)+\gamma]/\Gamma^{\rm total}$ ($n\ge
n'$) for these decays. Thus for this comparison it is necessary to get
theoretical predictions for the total decay rates of $\chi_b(nP)$. The
main decay channels of the bottomonium $P$ states are inclusive strong
decays to gluon and quark states and radiative decays. The
strong decays were extensively studied in the literature
\cite{had,kmrr} including leading-order QCD corrections. The two-gluon
annihilation rates of ${}^3\!P_0$  states with the
account of relativistic corrections are given in our model
\cite{fgq88}  by

\begin{eqnarray}
  \label{eq:2gp0}
  \Gamma({}^3\!P_0\to gg)&=&\frac{8\alpha_s^2}{3M^2}\Biggl|\int
  \frac{d^3p}{(2\pi)^3} \frac{m_Q^2}{Mp}\Biggl[\frac{p}{\epsilon_Q(p)}
  \ln\frac{\epsilon_Q(p)+p}{\epsilon_Q(p)-p}\cr\nonumber\\
&&  +\left(1-\frac{M}{2\epsilon_Q(p)}\right)
\left(2- \frac{\epsilon_Q(p)}{p}
\ln\frac{\epsilon_Q(p)+p}{\epsilon_Q(p)-p}\right)\Biggr]\phi_P(p)\Biggr|^2.
 \end{eqnarray}
To include  leading-order QCD corrections  \cite{kmrr} this expression
should be multiplied by  $(1+10.0\alpha_s/\pi)$  
for $\chi_{b0}$ and $(1+10.2\alpha_s/\pi)$ for $\chi_{b0}'$. The
corresponding expression for two-gluon annihilation rates of
${}^3\!P_2$ states reads as \cite{fgq88} 
\begin{eqnarray}
  \label{eq:2gp2}
 \Gamma({}^3\!P_2\to gg)&=&\frac{8\alpha_s^2}{5M^2}\Biggl\{\Biggl|\int
  \frac{d^3p}{(2\pi)^3} \frac{m_Q\epsilon_Q(p)}{Mp}\Biggl[\left(2+
\frac{p}{\epsilon_Q(p)}
  \ln\frac{\epsilon_Q(p)+p}{\epsilon_Q(p)-p}
-\frac{\epsilon_Q(p)}{p}\ln\frac{\epsilon_Q(p)+p}{\epsilon_Q(p)-p}
\right)
\cr \nonumber\\&& \times
\left(1+\frac{m_Q}{2\epsilon_Q(p)[\epsilon_Q(p)+m_Q]}\right)-\frac{2p^2}
{3\epsilon_Q(p)[\epsilon_Q(p)+m_Q]}\Biggr] \phi_P(p)\Biggr|^2 
\cr \nonumber\\&&+\frac16 \Biggl|\int
  \frac{d^3p}{(2\pi)^3} \frac{m_Q^2}{Mp}\Biggl[\frac{p}{\epsilon_Q(p)}
  \ln\frac{\epsilon_Q(p)+p}{\epsilon_Q(p)-p}
+3\left(2-\frac{\epsilon_Q(p)}{p}\ln\frac{\epsilon_Q(p)+p}
{\epsilon_Q(p)-p}\right)\cr\nonumber\\
&&-\frac{\epsilon_Q(p)}{m_Q}\left(2-\frac{M}{m_Q}
\right)\left(2-\frac{\epsilon_Q(p)}{p}\ln\frac{\epsilon_Q(p)+p}
{\epsilon_Q(p)-p}\right)
+\frac{p^2}{m_Q[\epsilon_Q(p)+m_Q]}\left(2-\frac{M}{m_Q}\right)\cr\nonumber\\
&&\times
\Bigglb(\frac12\left(\frac{\epsilon_Q(p)}{p}\ln\frac{\epsilon_Q(p)+p}
{\epsilon_Q(p)-p}-2\right)\left(1-\frac{3\epsilon_Q^2(p)}{p^2}\right)
+1\Biggrb)\Biggr]\phi_P(p)\Biggr|^2\Biggr\},
  \end{eqnarray}
with leading-order QCD corrections \cite{kmrr} accounted by the
factors $(1-0.1\alpha_s/\pi)$ 
for $\chi_{b2}$ and $(1+1.0\alpha_s/\pi)$ for $\chi_{b2}'$. Here $\phi_P(p)$
is the $P$ state radial wave function in momentum space. In the
nonrelativistic limit $p/m_Q\to 0$ and $M\to 2m_Q$,
Eqs.~(\ref{eq:2gp0}) and (\ref{eq:2gp2}) reduce to the known
expressions \cite{kmrr}:
\begin{eqnarray}
  \label{eq:2gnr}
\Gamma({}^3\!P_0\to gg)&=&\frac{6\alpha_s^2\left|R'_{nP}(0)\right|^2}
{m_Q^4},\\
\Gamma({}^3\!P_2\to gg)&=&\frac{8\alpha_s^2\left|R'_{nP}(0)\right|^2}
{5m_Q^4}.
\end{eqnarray}
The calculation of radiative and relativistic corrections to the
annihilation decay rates of ${}^3\!P_1$ and ${}^1\!P_1$ states is a very
complicated problem which has not been solved yet. Thus to estimate
their hadronic decay rates we use the tree-level nonrelativistic
expressions \cite{kmrr}
\begin{eqnarray}
  \label{eq:p1g}
  \Gamma({}^3\!P_1\to q\bar q+g)&=&\frac{8\alpha_s^3n_f}{9\pi m_Q^4}
  \left|R'_{nP}(0)\right|^2\ln(m_Q\left<r\right>),\\
\Gamma({}^1\!P_1\to ggg)&=&\frac{20\alpha_s^3}{9\pi m_Q^4}
  \left|R'_{nP}(0)\right|^2\ln(m_Q\left<r\right>), \\
\Gamma({}^1\!P_1\to
  gg+\gamma)&=&\frac{36}5e_q^2\frac{\alpha}{\alpha_s} 
\Gamma({}^1\!P_1\to ggg).
\end{eqnarray}
From the relativistic consideration of the decays of ${}^3\!P_{0,2}$
states, which shows that relativistic effects give $\sim 10\%$
contributions to the $b\bar b$ decay rates, we can expect that these
formulae give a reasonable estimate of the corresponding decay rates. 
For numerical calculations of hadronic decay rates of $\chi_b$ states
we use $\alpha_s=0.18$ obtained from the experimental ratio of
$\Gamma(\Upsilon\to gg\gamma)/\Gamma(\Upsilon\to ggg)$ \cite{gr}. The
calculated partial decay rates and branching fractions for $1P$ and $2P$
states of the bottomonium are compared with available experimental
data in Table~\ref{tab:1p2pb}. There we give both PDG \cite{pdg}
averages and very recent CLEO \cite{cleo} data. 
We see that our predictions for the
branching fractions for radiative transitions of $\chi_b(nP)$ to
$\Upsilon(n'S)$ states are in good agreement with experiment. The only
discrepancy (of $2.5\sigma$) is the CLEO value for 
${\cal  B}(\chi_{b1}(2P)\to \Upsilon(2S) +\gamma)$ transition which is
approximately two times larger than our model prediction and PDG value.  
The CLEO Collaboration \cite{cleo} measured also two photon decays of
$\Upsilon(3S)$ via $\chi_b(1P_J)$ states. They report the branching
fraction for $\Upsilon(3S)\to \chi_b(1P_J)+\gamma\to \Upsilon(1S)+
\gamma\gamma$ transitions summed over all the $J$ states:
\begin{equation}\label{cleobr}
{\cal B}(\Upsilon(3S)\to \Upsilon(1S)+\gamma\gamma)=(2.14\pm 0.22\pm
0.21) \times 10^{-3}.
\end{equation}
Using our model results in Tables~\ref{tab:e1b}--\ref{tab:1p2pb} for
corresponding decay rates, we get
\begin{eqnarray*}
 {\cal B}(\Upsilon(3S)\to \chi_{b0}(1P)+\gamma \to
 \Upsilon(1S)+\gamma\gamma)&=&1.4 \times 10^{-5},\cr
{\cal B}(\Upsilon(3S)\to \chi_{b1}(1P)+\gamma \to
 \Upsilon(1S)+\gamma\gamma)&=&9.97 \times 10^{-4},\cr
{\cal B}(\Upsilon(3S)\to \chi_{12}(2P)+\gamma \to
 \Upsilon(1S)+\gamma\gamma)&=&9.96 \times 10^{-4},  
\end{eqnarray*}
and the sum over all the $J$ states is equal to
\[
{\cal B}(\Upsilon(3S)\to \Upsilon(1S)+\gamma\gamma)=2.04 \times
10^{-3} 
\]
in accord with CLEO data (\ref{cleobr}).

\begin{table}[htb]
  \caption{Partial decay rates and branching fractions for $1P$ and
    $2P$ states of bottomonium.}  
  \label{tab:1p2pb}
\begin{ruledtabular}
\begin{tabular}{cccccc}
Level& Decay & $\Gamma$ (keV) & $\cal B$ (\%)
&\multicolumn{2}{c}{$\cal B^{\rm exp}$ (\%)}\\
& & & &PDG \cite{pdg} & CLEO \cite{cleo}\\ 
\hline
$1\,{}^3\! P_0$ & $gg$ & 653 & 95.6 &  &\\
          &$1\,{}^3\!S_1+\gamma$ & 29.9 & 4.4 & $< 6$ &\\
$1\,{}^3\! P_1$ & $q\bar q+g$ & 57 & 60.9 & \\
          &$1\,{}^3\!S_1+\gamma$ & 36.6 & 39.1 & $35\pm8$ & \\
$1\,{}^3\! P_2$ & $gg$ & 109 & 73 &\\
          &$1\,{}^3\!S_1+\gamma$ & 40.2 & 27& $22\pm 4$& \\
$1\,{}^1\! P_1$ & $ggg$ & 36 & 40.1 & &\\
                & $gg+\gamma$ & 1.2 & 1.3 &\\
          &$1\,{}^1\!S_0+\gamma$ & 52.6 & 58.6& &\\
$2\,{}^3\! P_0$ & $gg$ & 431 & 95.8 & &\\
          &$2\,{}^3\!S_1+\gamma$ & 11.0 & 2.4 & $4.6\pm2.1$ &
                $2.59\pm0.92\pm 0.51$\\
          &$1\,{}^3\!S_1+\gamma$ & 6.8 & 1.5 & $0.9\pm0.6$ & $<1.44$\\ 
          &$1\,{}^3\!D_1+\gamma$ & 1.17 & 0.3 & & \\
$2\,{}^3\! P_1$ & $q\bar q+g$ & 50 & 67.2 & & \\
          &$2\,{}^3\!S_1+\gamma$ & 14.7 & 19.8 & $21\pm4$ & $41.5\pm
          1.2\pm 5.9$\\
          &$1\,{}^3\!S_1+\gamma$ & 7.5 & 10.0 & $8.5\pm1.3$ & $11.6\pm
          0.4\pm 0.9$\\
          &$1\,{}^3\!D_1+\gamma$ & 0.6 & 0.8 &  &\\
          &$1\,{}^3\!D_2+\gamma$ & 1.6 & 2.2 &  &\\
$2\,{}^3\! P_2$ & $gg$ & 76& 73.4 & & \\
          &$2\,{}^3\!S_1+\gamma$ & 16.7 & 16.1 & $16.2\pm2.4$&
          $19.3\pm 1.1\pm3.1$\\
          &$1\,{}^3\!S_1+\gamma$ & 8.0 & 7.7& $7.1\pm 1.0$&
          $7.0\pm0.4\pm0.8$ \\
          &$1\,{}^3\!D_1+\gamma$ & 0.04 & 0.03 & &\\
          &$1\,{}^3\!D_2+\gamma$ & 0.5 & 0.5 & & \\ 
          &$1\,{}^3\!D_3+\gamma$ & 2.4 & 2.3 & & \\
$2\,{}^1\! P_1$ & $ggg$ & 31.5 & 47.9 & &\\
                & $gg+\gamma$ & 1.1 & 1.7 & &\\
          &$2\,{}^1\!S_0+\gamma$ & 21.4 & 32.5& &\\
          &$1\,{}^1\!S_0+\gamma$ & 9.4 & 14.3& & \\
          &$1\,{}^1\!D_2+\gamma$ & 2.4 & 3.6 & &
\end{tabular}
\end{ruledtabular}
\end{table}

The CLEO Collaboration \cite{cleo} presented recently the first
evidence for the production of the triplet $\Upsilon(1D)$ states in
the four photon transitions $3S\to 2P+\gamma\to 1D+\gamma\gamma\to
1P+\gamma\gamma\gamma \to 1S+\gamma\gamma\gamma\gamma \to
e^+e^-+\gamma\gamma\gamma\gamma$. The measured product branching
fraction for these five decays is equal to $(3.3\pm 0.6 \pm 0.5)\times
10^{-5}$. In Table~\ref{tab:4gamma} we give the theoretical
predictions for the branching fractions of such four photon decays in
our model and in the recent quark model analysis of Godfrey and Rosner
\cite{grD}. In general, both theoretical predictions are consistent. In
the analysis of Ref.~\cite{grD}, the dominant decay is  $3\,{}^3\!S_1\to
2\,{}^3\!P_1+\gamma\to 1\,{}^3\!D_2+\gamma\gamma\to 1\,{}^3\!P_1
+\gamma\gamma\gamma \to 1\,{}^3\!S_1+\gamma\gamma\gamma\gamma\to
e^+e^-+\gamma\gamma\gamma\gamma$, while
in our model the above and $3\,{}^3\!S_1\to
2\,{}^3\!P_2+\gamma\to 1\,{}^3\!D_3+\gamma\gamma\to 1\,{}^3\!P_2
+\gamma\gamma\gamma \to 1\,{}^3\!S_1+\gamma\gamma\gamma\gamma\to
e^+e^-+\gamma\gamma\gamma\gamma$
transitions have almost the same rate and dominate. In the last line
we give the sum of all these decay channels. Both theoretical
predictions agree with CLEO measurement.        

\begin{table}[htb]
   \caption{Predicted branching fractions of four photon decays of
$\Upsilon(3S)$ involving $1D$ states  corresponding to $3S\to
2P+\gamma\to 1D+\gamma\gamma\to 1P+\gamma\gamma\gamma \to
1S+\gamma\gamma\gamma\gamma \to e^+e^-+\gamma\gamma\gamma\gamma$
transition. The last line gives the sum of all decay channels.} 
  \label{tab:4gamma}
\begin{ruledtabular}
 \begin{tabular}{ccccc}
$2\,{}^3\! P_J$ state&$1\,{}^3\! D_J$ state & $1\,{}^3\! P_J$ state &
${\cal B}^{\rm our}$ & ${\cal B}$ \cite{grD} \\
 & & &  $(\times 10^{-6})$ & $(\times 10^{-6})$\\
\hline
$2\,{}^3\! P_2$ & $1\,{}^3\! D_3$& $1\,{}^3\! P_2$ & 15.2 & 7.8 \\
                & $1\,{}^3\! D_2$& $1\,{}^3\! P_2$ & 0.7 &0.3 \\
                &                & $1\,{}^3\! P_1$ & 3.9 &2.7 \\
                & $1\,{}^3\! D_1$& $1\,{}^3\! P_2$ & 0.0 &0.0 \\
                &                & $1\,{}^3\! P_1$ & 0.1 &0.1 \\
                 &                & $1\,{}^3\! P_0$ & 0.0 &0.0 \\
$2\,{}^3\! P_1$ & $1\,{}^3\! D_2$& $1\,{}^3\! P_2$ & 2.9 & 2.5 \\
                &                & $1\,{}^3\! P_1$ & 15.5 &20.1 \\
                & $1\,{}^3\! D_1$& $1\,{}^3\! P_2$ & 0.1 &0.1 \\
                &                & $1\,{}^3\! P_1$ & 2.4 &3.3 \\
                 &                & $1\,{}^3\! P_0$ & 0.5 &0.4 \\
$2\,{}^3\! P_0$ & $1\,{}^3\! D_1$& $1\,{}^3\! P_2$ & 0.0 & 0.0 \\
                &                & $1\,{}^3\! P_1$ & 0.5 &0.3 \\
                 &                & $1\,{}^3\! P_0$ & 0.0 &0.0 \\
all & all & all & 41.8 & 37.6
  \end{tabular}
\end{ruledtabular}
\end{table}

An important problem of quarkonium physics is the search for
bottomonium spin singlet states $\eta_b(n\,{}^1\!S_0) $ and
$h_b(n\,{}^1\!P_1)$. In the previous section we discussed the
possibilities of finding $\eta_b$ in radiative M1 decays. From
Table~\ref{tab:1p2pb} we see that these two states can be discovered
simultaneously. Indeed the radiative decay
$h_b(1\,{}^1\!P_1) \to \eta_b(1\,{}^1\!S_0)+\gamma$ with the photon
energy of 480 MeV is the main decay channel of $h_b$ (the branching
fraction of this decay exceeds 50\%). Thus production of a few $h_b$
states, e.g., through $\Upsilon''(3S)\to h_b(1\,{}^1\!P_1)\pi\pi$ or
$\Upsilon''(3S)\to h_b(1\,{}^1\!P_1)\pi$
decays, which branching fractions are predicted to be about
0.1--1\% \cite{kyv}, will give a good possibility of finding $\eta_b$.

\begin{table}[htb]
  \caption{Radiative E1 transition rates of the $B_c$ meson.}
  \label{tab:e1bc}
\begin{ruledtabular}
\begin{tabular}{ccccccccc}
Decay& $\omega$ & $\Gamma^{\rm NR}$ & $\Gamma^V$ & $\Gamma^S$ & $\Gamma$ &
$\Gamma$ \cite{eq} & $\Gamma$ \cite{gklt} & $\Gamma$ \cite{ful}     \\
     & MeV & keV & keV & keV & keV & keV & keV & keV\\
\hline
$1\,{}^3\! P_0\to1\,{}^3\! S_1\gamma$ & 355 & 75.5 & 74.1 & 81.3 & 67.2 & 79.2 &
65.3 & 74.2\\
$1P1\to1\,{}^3\! S_1\gamma$ & 389 & 87.1 & 82.9 & 87.1 & 78.9 & 99.5 & 77.8 &
75.8 \\
$1P1'\to1\,{}^3\! S_1\gamma$ & 405 & 13.7 & 12.6 & 11.6 & 13.6 & 0.1 & 8.1 &
26.2 \\
$1\,{}^3\! P_2\to1\,{}^3\! S_1\gamma$ & 416 & 122 & 105 & 102 & 107 & 112.6 & 102.9
& 126 \\
$1P1\to1\,{}^1\! S_0\gamma$ & 447 & 18.4 & 16.3 & 14.4 & 18.4 & 0.0 & 11.6 &
32.5 \\
$1P1'\to1\,{}^1\! S_0\gamma$ & 463 & 147 & 134 & 136 & 132 & 56.4 & 131.1 &
128 \\
$2\,{}^3\! S_1\to1\,{}^3\! P_0\gamma$ & 181 & 5.53 & 4.36 & 5.00 & 3.78 & 7.8 & 7.7
& 9.6 \\
$2\,{}^3\! S_1\to1P1\gamma$ & 146 & 7.65 & 5.98 & 6.98 & 5.05 & 14.5 & 12.8 &
13.3 \\
$2\,{}^3\! S_1\to1P1'\gamma$ & 130 & 0.74 & 0.62 & 0.61 & 0.63 & 0.0 & 1.0 &
2.5 \\
$2\,{}^3\! S_1\to1\,{}^3\! P_2\gamma$ & 118 & 7.59 & 5.99 & 6.86 & 5.18 & 17.7 &
14.8 & 14.5 \\
$2\,{}^1\! S_0\to1P1\gamma$ & 101 & 1.05 & 0.94 & 0.86 & 1.02 & 0.0 & 1.9 &
6.4 \\
$2\,{}^1\! S_0\to1P1'\gamma$ & 84 & 4.40 & 3.77 & 3.82 & 3.72 & 5.2 & 15.9 &
13.1 \\
$2\,{}^3\! P_0\to2\,{}^3\! S_1\gamma$ & 207 & 34.0 & 28.8 & 28.4 & 29.2 & 41.2 &
25.5 & \\
$2P1\to2\,{}^3S_1\gamma$ & 241 & 45.3 & 37.6 & 37.3 & 37.9 & 54.3 & 32.1 &
\\
$2P1'\to2\,{}^3\! S_1\gamma$ & 259 & 10.4 & 8.45 & 7.86 & 9.07 & 5.4 & 5.9 &
\\
$2\,{}^3\! P_2\to2\,{}^3\! S_1\gamma$ & 270 & 75.3 & 58.9 & 60.6 & 57.3 & 73.8 &
49.4 & \\
$2P1\to2\,{}^1\! S_0\gamma$ & 285 & 13.8 & 11.1 & 10.5 & 11.7 &  & 8.1 & \\
$2P1'\to2\,{}^1\! S_0\gamma$ & 303 & 90.5 & 73.2 & 73.8 & 72.5 &  & 58.0 & 
\end{tabular}
\end{ruledtabular}
\end{table}

\begin{table}[htb]
  \caption{Radiative E1 transition rates of the $B_c$ meson involving $D$
    states.} 
  \label{tab:e1bc2}
\begin{ruledtabular}
\begin{tabular}{cccccccc}
Decay& $\omega$ & $\Gamma^{\rm NR}$ & $\Gamma^V$ & $\Gamma^S$ & $\Gamma$ &
$\Gamma$ \cite{eq} & $\Gamma$ \cite{gklt}     \\
     & MeV & keV & keV & keV & keV & keV & keV \\
\hline
$2\,{}^3\! P_2\to1\,{}^3\! D_1\gamma$ & 84 & 0.035 & 0.031 & 0.027 & 0.035 & 0.2 &
0.1 \\
$2\,{}^3\! P_2\to1D2\gamma$ & 79 & 0.285 & 0.245 & 0.222 & 0.269 & 3.2 & 1.5
\\
$2\,{}^3\! P_2\to1D2'\gamma$ & 77 & 0.139 & 0.114 & 0.116 & 0.113 &  & 0.5 \\
$2\,{}^3\! P_2\to1\,{}^3\! D_3\gamma$ & 75 & 2.08 & 1.64 & 1.69 & 1.59 & 17.8 & 10.9
\\
$2\,{}^3\! P_0\to1\,{}^3\! D_1\gamma$ & 19 & 0.041 & 0.034 & 0.033 & 0.036 & 6.9 &
3.2 \\
$2P1\to1\,{}^3\! D_1\gamma$ & 54 & 0.204 & 0.174 & 0.165 & 0.184 & 0.3 & 1.6
\\
$2P1'\to1\,{}^3\! D_1\gamma$ & 73 & 0.070 & 0.062 & 0.052 & 0.073 & 0.4 & 0.3
\\
$2P1\to1D2\gamma$ & 49 & 0.517 & 0.420 & 0.422 & 0.418 & 9.8 & 3.9 \\
$2P1\to1D2'\gamma$ & 47 & 0.023 & 0.019 & 0.018 & 0.021 &  & 1.2 \\
$2P1'\to1D2\gamma$ & 68 & 0.172 & 0.142 & 0.135 & 0.149 & 11.5 & 2.5
\\
$2P1'\to1D2'\gamma$ & 66 & 1.49 & 1.20 & 1.20 & 1.20 &  & 3.5 \\
$1^3\! D_1\to1^3\! P_0\gamma$ & 365 & 133 & 119 & 110 & 128 & 88.6 & 79.7 \\
$1^3\! D_1\to1P1\gamma$ & 331 & 65.3 & 63.7 & 54.2 & 73.8 & 49.3 & 39.2
\\
$1\,{}^3\! D_1\to1P1'\gamma$ & 315 & 7.81 & 6.91 & 6.20 & 7.66 & 0.0 & 3.3 \\
$1\,{}^3\! D_1\to1\,{}^3\! P_2\gamma$ & 303 & 3.82 & 4.27 & 3.17 & 5.52 & 2.7 & 2.2
\\
$1D2\to1P1\gamma$ & 335 & 139 & 112 & 113 & 112& 88.8 & 44.6 \\
$1D2\to1P1'\gamma$ & 319 & 14.9 & 13.4 & 12.7 & 14.1 & 0.1 & 18.4 \\
$1D2'\to1P1\gamma$ & 338 & 7.10 & 6.70 & 6.17 & 7.25 & & 25.0 \\
$1D2'\to1P1'\gamma$ & 321 & 143 & 116 & 117 & 116 & 92.5 & 46.0 \\
$1D2\to1\,{}^3\! P_2\gamma$ & 308 & 23.6 & 23.4 & 19.7 & 27.5 & 24.7 & 12.2
\\
$1D2'\to1\,{}^3\! P_2\gamma$ & 310 & 12.6 & 11.4 & 10.1 & 12.8 & & 6.8 \\
$1\,{}^3\! D_3\to1\,{}^3\! P_2\gamma$ & 312 & 149 & 112 & 122 & 102 & 98.7 & 76.9 
\end{tabular}
\end{ruledtabular}
\end{table}

In Tables~\ref{tab:e1bc} and \ref{tab:e1bc2} we compare our
results for the E1 radiative decay rates of the $B_c$ meson with other
quark model predictions \cite{eq,gklt,ful}. Comparison of the
calculations using relativistic $\Gamma$ and nonrelativistic
$\Gamma^{\rm NR}$ formulae for decay rates shows that relativistic
corrections do not exceed 20\% in $B_c$ meson E1 decays. Most of the
theoretical predictions for E1 transitions between $P$ and $S$ states
of $B_c$ mesons given in Table~\ref{tab:e1bc} are compatible with each
other. The largest 
differences occur for decays involving $P1$ and $P1'$ states which are
the mixtures of spin singlet and spin triplet states (\ref{eq:pmix})
due to different mixing angles used by the authors. Note that for such
transitions there are additional relativistic spin-flip contributions
to decay rates (\ref{eq:edr}) proportional to $\chi(J',L',J,L)$
(\ref{eq:chi}) which are specific only for $B_c$ mesons. In general, our
predictions are closer to the ones of Ref.~\cite{gklt}.

In Table~\ref{tab:e1bc2} we present the E1 radiative decay rates of
$B_c$ mesons where either the initial or final state is a $D$ wave
state. Here we find rather large variations in theoretical
predictions. The main reason of these distinctions is the difference
in values of $D$ state masses, which for some states reaches
70~MeV (see Table~\ref{tab:bcm}). Since $2P$ and $1D$ states of the $B_c$
are rather close, such a difference significantly influences the energy of
the emitted photon and thus the decay rates. For decays involving the
mixed spin singlet and spin triplet states $P1$, $P1'$ (\ref{eq:pmix})
and $D2$, $D2'$ (\ref{eq:dmix}) the additional relativistic spin-flip
contributions (\ref{eq:edr}) are important, especially for transitions
where both initial and final states are  mixed states.

\section{Conclusions}
\label{sec:conc}

In this paper the mass spectra and radiative M1 and E1 decay rates of
charmonium, bottomonium and $B_c$ mesons were calculated in the
relativistic quark model based on the quasipotential approach in
quantum field theory. Special attention was  devoted to the
role of the relativistic effects in these processes. Since both quarks
in the considered mesons are heavy, the $v/c$ expansion was applied. In
the mass spectra calculations retardation as well as one-loop radiative
corrections were taken into account. We also included the one-loop
correction due to the finite charm quark mass to the bottomonium mass
spectrum. It was found that this correction is rather small \cite{efgmc}
and its inclusion allows one to obtain an even better fit of the
bottomonium  excited states with the slightly shifted value of QCD
parameter $\Lambda$. The calculated mass spectra of charmonium and bottomonium
agree with the experimental data within a few MeV. Comparison of our
results for the $B_c$ meson mass spectrum with previous calculations
showed that different predictions for ground state and low excitations
agree within 30~MeV.

The pseudoscalar and vector decay constants of $B_c$ meson were
calculated using the relativistic wave functions obtained during the
mass spectrum calculations. It was found that relativistic effects
reduce these constants by approximately 20\% and produce the splitting
between them of about 70~MeV.   

It was shown that relativistic effects play a significant role in
radiative decays of mesons. Their form strongly depends on the
Lorentz structure of the quark-antiquark interaction. The most
sensitive are radiative M1 decays, where even for allowed transitions
they significantly influence predictions for the rates. An
important example is the $J/\Psi\to \eta_c\gamma$ transition, which is
overestimated by a factor of more than two if the nonrelativistic
approximation is used. It is argued that the inclusion of relativistic
corrections for a pure scalar or vector confining potential is not
enough to bring theoretical predictions in accord with
experiment. Only for the specific mixture of these potentials (\ref{vlin})
with the mixing coefficient $\varepsilon=-1$, the agreement can be
obtained. For other decay rates this mixing of scalar and vector
potentials also gives the best results. The hindered M1 transition rates
are dominated by relativistic contributions and are significantly
enhanced by them. The comparison of the allowed and hindered M1 rates
in bottomonium shows that the latter provide better opportunity for
the search of the missing pseudoscalar $\eta_b$ state of the bottomonium.

The analysis of radiative E1 transitions showed that the form of
relativistic corrections is less dependent on the Lorentz structure of
the quark potential than in the case of M1 transitions. However, for some
decays, e.g., $\Psi'\to \chi_{cJ}\gamma$, the consideration of the
mixed (\ref{vlin})  vector and scalar potentials (with the same value
of $\varepsilon=-1$) is important for bringing decay rates in accord
with experimental data. In general, all our predictions for radiative
decay rates and branching fractions of charmonium and bottomonium
agree with measured values. In the case of the $B_c$ meson radiative E1
decays an important additional relativistic correction to decay rates which
causes the flip of the quark spin was found. This contribution
(\ref{eq:chi}) to the radiative E1 decay rate  (\ref{eq:edr}) is
specific only for transitions involving mixed states  $nP1$, $nP1'$
(\ref{eq:pmix}) or $nD2$, $nD2'$ (\ref{eq:dmix}) of $B_c$ and is
caused by the difference of the $c$ and $b$ quark masses. Finally, a
comparison of various quark model predictions for the radiative M1
and E1 decay rates of $B_c$ has been performed. These radiative 
transitions along with pionic ones are the main decay channels of the
low lying excitations in the $B_c$ meson.               

\acknowledgments
The authors express their gratitude to A. Martin, M. M\"uller-Preussker and
V. Savrin for support and discussions. We are grateful to J. L. Rosner
for stimulating correspondence and quoting some of our preliminary
results on bottomonium radiative transitions  in Ref.~\cite{grnb}.
Two of us (R.N.F and V.O.G.) were supported in part by the 
{\it Deutsche Forschungsgemeinschaft} under contract Eb 139/2-1,
{\it Russian Foundation for Basic Research} under Grant
No.~00-02-17768 and {\it Russian Ministry
  of Education} under Grant No.~E00-3.3-45.

\end{document}